\documentclass[gmd, manuscript]{copernicus}

\begin{document}

\nolinenumbers

\title{Evaluating extreme precipitation forecasts: A threshold-weighted, spatial verification approach for comparing an AI weather prediction model against a high-resolution NWP model}


\Author[1][nicholas.loveday@bom.gov.au]{Nicholas}{Loveday} 
\Author[2][]{Tracy}{Hertneky}

\affil[1]{Bureau of Meteorology, Melbourne, Australia}
\affil[2]{NSF NCAR and DTC, Boulder, USA}




\runningtitle{TEXT}

\runningauthor{TEXT}

\received{}
\pubdiscuss{} 
\revised{}
\accepted{}
\published{}


\firstpage{1}

\maketitle

\begin{abstract}
Recent advances in AI-based weather prediction have led to the development of artificial intelligence weather prediction (AIWP) models with competitive forecast skill compared to traditional NWP models, but with substantially reduced computational cost. 
There is a strong need for appropriate methods to evaluate their ability to predict extreme weather events, particularly when spatial coherence is important, and grid resolutions differ between models.

We introduce a verification framework that combines spatial verification methods and weighted proper scoring rules. 
Specifically, the framework extends the High-Resolution Assessment (HiRA) approach with threshold-weighted scoring rules. It enables user-oriented evaluation consistent with how forecasts may be interpreted by operational meteorologists or used in simple post-processing systems.
The method supports targeted evaluation of extreme events by allowing flexible weighting of the relative importance of different decision thresholds.
We demonstrate this framework by evaluating 32 months of precipitation forecasts from an AIWP model and a high-resolution NWP model.
Our results show that model rankings are sensitive to the choice of neighbourhood size. Increasing the neighbourhood size has a greater impact on scores evaluating extreme-event performance for the high-resolution NWP model than for the AIWP model.
At near equivalent neighbourhood sizes, the empirical CDF of the high-resolution NWP model only outperformed the empirical CDF of the AIWP model in predicting extreme precipitation events at short lead times.
We also demonstrate how this approach can be extended to evaluate discrimination ability in predicting heavy precipitation. We find that the high-resolution NWP model had superior discrimination ability at short lead times. 
\end{abstract}


\introduction  
In recent years, there has been rapid progress in the development of artificial intelligence weather prediction (AIWP) models, sometimes referred to as data-driven models \citep{Ben_Bouall_gue_2024}. These models are typically based on neural networks trained on reanalysis datasets. This approach contrasts with traditional numerical weather prediction (NWP) models, which rely on equations that model physical processes. One major advantage of AIWP models is that they are significantly more computationally efficient than NWP models for making predictions once they have been trained.

Most early global AIWP models are trained on ERA5 \citep{hersbach2020era5} data, and therefore have a similar or coarser grid resolution of approximately $0.25^{\circ}$. These include deterministic models \citep[e.g.,][]{keisler2022forecastingglobalweathergraph, Lam_2023, Bi_2023, lang2024aifsecmwfsdatadriven} and ensemble models \citep[e.g.,][]{Price_2024, Lang_2026}. In general, these models operate at much coarser resolutions than the high-resolution physical NWP models available to many international weather forecasting centers. However, higher-resolution, limited-area AIWP models have started to emerge more recently \citep[e.g.,][]{Nipen_2026, adamov2025buildingmachinelearninglimited, abdi2025hrrrcastdatadrivenemulatorregional}, and this trend is expected to continue.

As modelling centers have access to high-resolution NWP models and with the accelerating development of higher-resolution AIWP models, there is a strong need for appropriate methodologies to compare the performance of models with differing spatial resolutions. Such comparisons are important to help meteorological agencies make informed decisions regarding future modelling strategies and to assist operational meteorologists and forecast system developers in understanding each model's strengths and weaknesses.

This raises several key questions we aim to address.

First, most benchmarking efforts of global AIWP models to date have relied on point-to-point verification methods \citep[e.g.,][]{Rasp_2024}, with perhaps the exception of \citet{Radford_2025_comparison}, who applied an object-based verification approach. Point-to-point methods may be appropriate for evaluating forecasts at individual locations (e.g., for end-users accessing weather forecasts via mobile apps). However, meteorologists and forecast systems often use forecasts spatially, where spatial coherence and physical realism are important.

Point-to-point verification methods tend to favour smoother forecasts than are typically observed in gridded observations \citep{subich2025fixingdoublepenaltydatadriven}. When spatial structure is important, point-to-point metrics can suffer from the ``double penalty'' effect \citep{Ebert_2008}. This can occur when a forecast feature matches the observed shape and intensity but is slightly displaced in location, leading to it being penalised twice. It is penalised once for the false prediction at the forecast location, and once for the missed event at the observed location. This issue is particularly problematic for high-resolution models, which aim to represent fine-scale features. Lower-resolution models that fail to capture the feature at all may score better under point-to-point metrics, because they incur only a single penalty. Thus, while high-resolution NWP models may produce forecasts that appear more realistic to meteorologists, they may not perform better under standard point-to-point verification metrics \citep{Mass_2002}.

To address the limitations of point-based verification, a wide range of spatial verification methods have been developed over the past two and a half decades \citep{Gilleland_2009, Dorninger_2018}. These methods are typically grouped into five categories: neighbourhood methods, scale-separation methods, feature-based (or object-based) methods, field-deformation methods, and distance measures. In this paper, we focus on an approach within the neighbourhood methods category \citep{Ebert_2009, Schwartz_2017}. Neighbourhood methods are verification approaches that evaluate forecasts using values within a local neighbourhood surrounding the observation location.

Second, there is a pressing need to assess how well both AIWP and high-resolution NWP models predict extreme weather events. Much of the existing literature has focused on a limited number of case studies \citep[e.g.,][]{Charlton_Perez_2024, morisseau2025object}. While case studies are valuable, they may be prone to selection bias or may not generalise across more cases. Focusing solely on cases where extremes occur can lead to the ``forecaster's dilemma'' \citep{Lerch_2017}. The forecaster's dilemma arises when forecasters or modellers must choose between issuing honest forecasts and adjusting their forecasts to optimise a particular scoring rule (we refer to the latter as ``hedging''). Proper scores discourage hedging and avoid the ``forecaster's dilemma''. A scoring rule is proper if the expected score is optimised by producing a forecast that corresponds to the forecaster's true belief \citep{Winkler_1968, Gneiting_2007}.  

There are some recent examples of broader evaluation of extremes across many events. For example, \citet{Lam_2023} and \citet{Ben_Bouall_gue_2024} evaluated the discrimination ability of AIWP models to differentiate between climatological extreme temperature events and non-events. \citet{Olivetti_2024} assessed performance in forecasting temperature and wind extremes primarily by conditioning on observed extremes, but also included an evaluation using a threshold-weighted squared error \citep{Taggart_2021}. \citet{mcgovern2026extremeweatherbenchframework} recently introduced a benchmark that focuses on evaluating several models to predict extremes based on a comprehensive set of events and marginal events. \citet{Zhang_2026} evaluated several AIWP models and ECMWF's HRES model against observations that are more extreme than the observations in the 1979--2017 training period of the AIWP models. Some of these studies have taken efforts to avoid, or practically reduce the impact of the ``forecaster's dilemma'', although some of these approaches still suffer from this problem.

These efforts have been limited to point-to-point verification methods, and there is a need for more comprehensive verification methods that assess the performance of models predicting extremes when they are used in a user-oriented spatial framework.

The appropriate verification approach depends on how forecasts are used in practice. Weather models are used in a wide variety of ways. For example, model output could be used at each grid point to produce forecasts for a user's location (e.g., via a mobile weather application). A high-resolution ensemble could be used by an operational meteorologist to infer the likelihood of different modes of convection during the afternoon. The dynamic tropopause output of a deterministic model (or single ensemble member) can also be used by a decision-support meteorologist to construct a conceptual model of the atmosphere, aiding rapid responses to bespoke queries in an emergency response centre. These use cases require different evaluation approaches. In the first example, point-to-point evaluation using proper or consistent scoring rules may be appropriate. In the second, point-to-point verification would not capture the spatial realism of convection within the ensemble, necessitating some form of spatial processing. Similarly, in the third case, evaluating the spatial structure of the dynamic tropopause may require a spatial verification method distinct from that used in the second example. These examples illustrate that no single metric can assess all the diverse ways that weather models are used.

It is desirable that forecasts are probabilistic and evaluated using proper scoring rules\footnote{In some cases, only a single-valued forecast is required for the forecast service rather than a full predictive distribution. Consistent scoring functions can then be used to evaluate forecasts expressed through a directive in the form of a statistical functional \citep{Gneiting_2011_point}; for example, a 90th percentile forecast can be evaluated using a quantile loss.} \citep{gneiting2014}. This is because probabilistic forecasts support optimal decision-making, and proper scores reward honest forecasts while discouraging hedging. Common examples of probabilistic forecasts in meteorology include ensembles, predictive distributions, or binary classifiers (i.e., probability forecasts for binary events). These forecasts are traditionally evaluated point-to-point; however, spatial processing can also be applied to construct probabilistic forecasts. For example, the high-resolution ensemble output in the second example could be used to create a probability mass function of various modes of convection, which could then be evaluated using a proper scoring rule. 

There have been at least two recent examples of using proper scoring rules to compare single-valued AIWP models against traditional NWP models. First, \citet{Brenowitz_2025} created a ``lagged'' ensemble of single-valued AIWP forecasts that could be evaluated with the CRPS. Secondly, \citet{gneiting2025probabilisticmeasuresaffordfair} converted single-valued forecasts into a predictive distribution using Isotonic Distributional Regression (IDR) \citep{Henzi_2021} and evaluated the resulting distribution using the CRPS. Both approaches are point-to-point methods that do not account for any spatial information.

One practical use case for weather models is when meteorologists visually assess a deterministic precipitation forecast around a point to qualitatively infer the likelihood of different precipitation amounts. A quantitative forecast can be constructed by generating an empirical cumulative distribution function (CDF) from a neighbourhood pseudo-ensemble around the observation. Similarly, post-processing techniques often employ neighbourhood approaches on NWP output to generate probabilistic forecasts \citep{Theis_2005, Schwartz_2017}. In this paper, we demonstrate a verification approach that merges two existing methods and assesses the model in one particular framework that aligns with a specific way that operational meteorologists may use a model. That is, we evaluate how well extremes are predicted when the model's spatial neighbourhood around a point is treated as an empirical CDF. The approach has several strengths including:
\begin{enumerate}
    \item It is a spatial verification method that aligns with a particular way that a model may be used.
    \item It uses proper scoring rules within a user-oriented framework.
    \item Threshold-weighting of proper scoring rules allows for the evaluation of extremes and other important decision thresholds.
    \item It permits each forecast system to be evaluated on its native grid when the forecast is defined as the finite neighbourhood empirical distribution.
    \item Models can be evaluated against station-based observations which may be desirable in some cases.
    \item It can be extended to measure the discrimination ability.
\end{enumerate}

\citet{Pagano_2024} raised the research question, ``Can spatial verification methods be extended to emphasize predictive performance for extremes without creating forecaster's dilemmas...?'' The approach presented here provides one response to this question by evaluating extremes within a user-oriented spatial framework that uses proper scoring rules. Specifically, we compare the native-grid neighbourhood forecast products derived from GraphCast-GFS and HRRR over 32 months. This comparison includes, rather than removes, differences associated with the grid spacing and effective resolution of the two forecast systems.
\section{Data}
\subsection{Observations}
Station data were used to assess the performance of the forecast models using a spatial verification approach. Most evaluation of AIWP models has been done against ERA5, which has some biases in its precipitation fields, with the United States having a dry bias \citep{Lavers_2022}. Additionally, most global AIWP models are trained against ERA5, and biases in ERA5 may propagate into these models and may go undetected if evaluated against ERA5. Some countries have alternative gridded rainfall datasets, which may serve as suitable substitutes, though these are not available everywhere. For this reason, some countries may place greater trust in verification using weather station data.

\begin{figure}[htbp]
    \centering
    \includegraphics[width=0.7\textwidth]{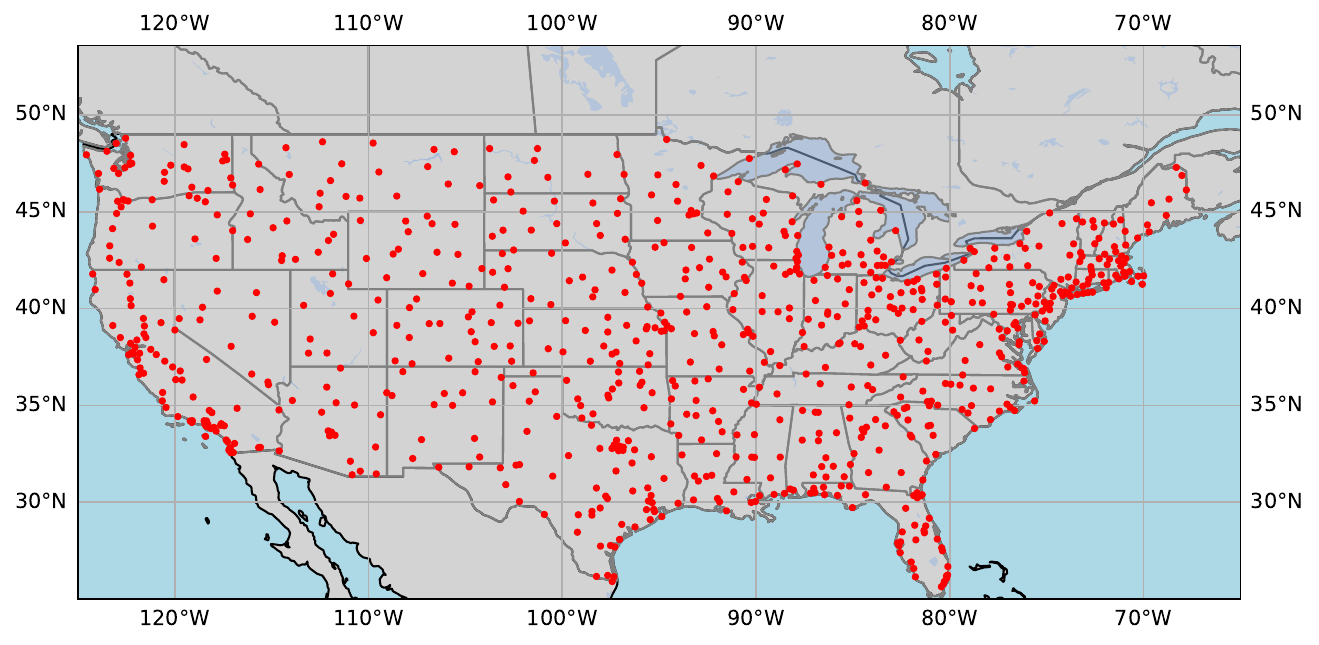}
    \caption{A map of ASOS station locations used in this study.}
    \label{fig:station_data}
  \end{figure}

While most past efforts have evaluated these models against gridded analysis, there have been limited examples of verification of AIWP models against station data \citep[e.g.,][]{Ben_Bouall_gue_2026}. WeatherReal \citep{jin2024} was recently developed to evaluate AIWP models against in situ observations. One limitation of this dataset is that it is only available for 2023, and our aim was to evaluate model performance over a longer period to assess extreme event prediction. Instead, one-minute Automated Surface Observing System (ASOS) \citep{NWS_ASOS} precipitation data were retrieved for the contiguous United States (CONUS). A map of ASOS weather station data is shown in Fig.~\ref{fig:station_data}. From this data, six-hour precipitation accumulations were derived. As a basic quality control measure, data were removed when one-minute accumulations exceeded 38~mm or six-hour accumulations exceeded 840 mm, which correspond to world record values \citep{WMO1994}. Additionally, six-hour accumulations were treated as missing when fewer than five hours of valid data existed within the six-hour period. ASOS data have previously been used to evaluate the HRRR model \citep[e.g.,][]{Ikeda_2013, Fovell_2022}, but, to our knowledge, have not been used to evaluate an AIWP model.

To define thresholds for extreme precipitation events, several options can be considered, each addressing slightly different questions. These include using fixed thresholds across all stations, annual climatological thresholds for each station, or seasonally varying climatological thresholds at each station. In this study, we adopt annual climatological thresholds for each station, which aligns with approaches used by some meteorological agencies in operational warning services.
Specifically, we define extreme precipitation events using the 99th and 99.9th percentile thresholds of six-hour precipitation accumulations at each station. As long observational time series are not available for all stations in the ASOS dataset, these thresholds are derived from ERA5 reanalysis data (1990–2020) at the grid point corresponding to each station location. These thresholds may differ from those derived directly from station observations; however, in the absence of a universally defined threshold for extreme precipitation, the exact choice of threshold is inherently somewhat subjective.

\subsection{Forecasts}
We used the 00 UTC run of two different models, with base-run times spanning from 1 January 2022 to 30 August 2024 (973 model runs).

The first model we evaluated was GraphCast, with initial conditions from NOAA's Global Forecast System (GFS), using the reforecast archive generated by \citet{Radford_2025}, hereafter referred to as GraphCast-GFS. The grid resolution of GraphCast-GFS is $0.25^{\circ}$. \citet{Radford_2025_comparison} found that, while GraphCast-GFS tended to overestimate lower rainfall amounts compared to GraphCast initialised with ECMWF’s Integrated Forecasting System (GraphCast-IFS), its performance for higher rainfall amounts was similar. GraphCast-GFS was selected because it was the only AIWP model output available at the time of analysis that produced six-hour precipitation accumulations over a sufficiently long period to assess performance in predicting extremes. 

Our evaluation begins in 2022 because the operational version of GraphCast was fine-tuned on data through 2021, and we sought to avoid testing the model on data used in its development. Since AIWP models can produce negative precipitation values, all negative values were set to zero, as this is a trivial post-processing step.

To compare an AIWP model with a high-resolution physical numerical weather prediction (NWP) model, we also evaluated version 4 of the High-Resolution Rapid Refresh (HRRR) model \citep{Dowell_2022}. HRRR is a convection-allowing, cloud-resolving physical NWP model with a horizontal grid spacing of 3 km, which runs hourly. HRRR was selected due to its widespread use in the United States as a high-resolution operational model.

Prior to score calculation, valid cases were matched between GraphCast-GFS and HRRR so that every paired comparison used the same observation times and stations. Since HRRR produces forecasts only up to 48 hours from the 00 UTC run, we limited the GraphCast-GFS evaluation to its first 48 forecast hours as well.

\section{Using HiRA and twCRPS}
This work unifies two different verification methods. We give an overview of each method.
\subsection{HiRA}
One neighbourhood verification method that can be used with point observations is the High-Resolution Assessment (HiRA) framework \citep{Mittermaier_2014}. One of the motivations for the development of the HiRA framework was to evaluate models of varying resolutions against site-based observations while avoiding the double penalty effect within a specified distance. 

The HiRA framework can be summarised as follows: for a given point observation, several squares (or sometimes circles) of varying sizes are used to define neighbourhoods of forecast grid cells surrounding the observation site. Each neighbourhood can then be used to generate a pseudo-ensemble, where each grid point within the neighbourhood is considered to have an equal probability of occurring. The pseudo-ensemble can subsequently be evaluated using a probabilistic score, such as the Brier score \citep{BRIER_1950}, the ranked probability score (RPS) \citep{Epstein_1969}, or the continuous ranked probability score (CRPS) \citep{Matheson_1976}. 

\subsection{Threshold-Weighted Continuous Ranked Probability Score}
In the context of evaluating the performance of forecasts for extreme events, threshold-weighted proper scoring rules are particularly important for forecast system developers and forecasters, as they help to avoid the ``forecaster's dilemma''. A scoring rule is considered proper if the expected score is optimised when the forecaster issues a probability forecast that corresponds to their true belief. The continuous ranked probability score (CRPS) is a strictly proper scoring rule that can be expressed as the integral of the Brier score over all possible thresholds \citep{Matheson_1976, Gneiting_2007}. The CRPS for evaluating a cumulative distribution function $F$ is defined as
\begin{align}
    \text{CRPS}(F, y) &= \int_{-\infty}^{\infty} \left(F(z) - \mathbb{1}\{y \leq z\} \right)^2 \,\mathrm{d}z\label{eq:CRPS-Brier}\\
    &= \mathbb{E}_F|X - y| - \frac{1}{2}\mathbb{E}_F|X - X'|.
\end{align}

In the first expression, $y$ is the observation, $z$ is the decision threshold, and $\mathbb{1}\{y \leq z\}$ is the indicator function, which equals 1 if $y \leq z$ and 0 otherwise. In the second expression, $X$ and $X'$ are independent random variables with distribution $F$, and $\mathbb{E}_F$ is expectation with respect to the probability distribution $F$. Lemma 2.2 in \citet{Baringhaus_2004} and equation 17 in \citet{Sz_kely_2005} show that the two expressions above are equal. When used to evaluate an ensemble forecast $F_{\mathrm{ens}}$ as an empirical CDF, the CRPS can also be expressed as

\begin{equation}
    \mathrm{CRPS}(F_{\mathrm{ens}}, y) = \frac{1}{M} \sum_{m=1}^M |x_m - y| - \frac{1}{2M^2} \sum_{m=1}^M \sum_{j=1}^M |x_m - x_j|,
    \label{eq:CRPS-eNRG}
\end{equation}

\noindent where $M$ is the number of ensemble members, $x_m$ and $x_j$ are the forecasts of a single ensemble member. \citet{Ferro_2013} proposed the ``fair'' CRPS for evaluating an ensemble which is defined as
\begin{equation}
    \mathrm{CRPS}_{\mathrm{fair}}(F_{\mathrm{ens}}, y) = \frac{1}{M} \sum_{m=1}^M |x_m - y| - \frac{1}{2M(M-1)} \sum_{m=1}^M \sum_{j=1}^M |x_m - x_j|.
    \label{eq:CRPS-fair}
\end{equation}

The fair correction addresses the effect of estimating an underlying predictive distribution with a finite number of ensemble members.

Weighted scoring rules have been developed to evaluate forecasts with an emphasis on specific ranges of decision thresholds, such as those that lie in the extremes of a climatological distribution. These scoring rules have proven useful in several contexts for assessing the ability to predict extreme events. For example, \citet{Loveday_2024} demonstrated that the standard mean squared error (MSE) showed no statistically significant difference in temperature forecast skill between meteorologists and automated guidance. However, using a threshold-weighted MSE revealed that meteorologists performed better at predicting temperature extremes than automated guidance. Similarly, \citet{wessel2024improvingprobabilisticforecastsextreme} showed that threshold-weighted scores can be used as a loss function to improve the predictive performance for extremes. \citet{Allen_2023} illustrated how multivariate threshold-weighted scores can be applied to evaluate forecasts of rainfall accumulation over consecutive days. 

Among the wide variety of threshold-weighted scores, the threshold-weighted continuous ranked probability score (twCRPS) has been the most extensively used and is defined as

\begin{equation}
    \mathrm{twCRPS}(F, y) = \int_{-\infty}^{\infty} \left(F(z) - \mathbb{1}\{y \leq z\} \right)^2 w(z)\,\mathrm{d}z,
\end{equation}

\noindent where $w(z)$ is a non-negative weight function \citep{Gneiting_2011}. The choice of $w(z)$ depends on the user’s decision thresholds of interest. For example, if the aim is to evaluate a model’s ability to produce accurate forecasts for users whose decision thresholds exceed 40$^{\circ}$C, a suitable threshold weight function could be $w(z) = \mathbb{1}(z > 40)$. Another example might involve choosing $w(z)$ to focus on temperature ranges associated with aircraft icing, in which case the weights could vary smoothly to reflect the relative importance of different thresholds $z$. 

\begin{figure}[htbp]
    \centering
    \includegraphics[width=0.5\textwidth]{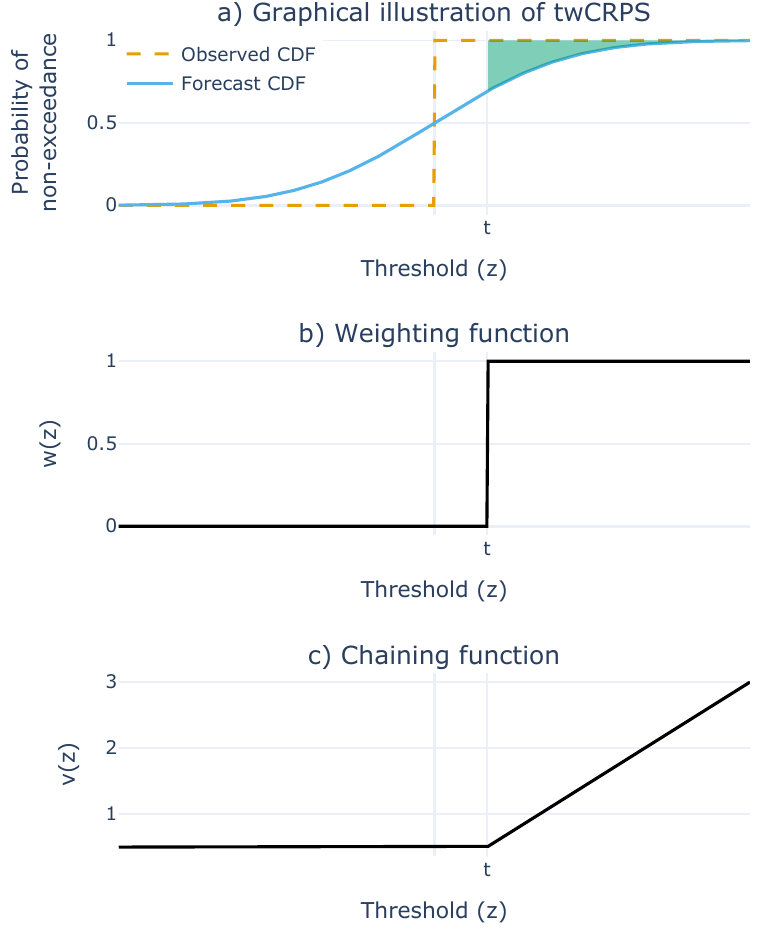}
    \caption{Graphical illustration of the threshold-weighted continuous ranked probability score (twCRPS) with a uniform weight of 1 applied to all thresholds $z > t$ and a weight of 0 applied elsewhere. \textbf{(a)} The solid blue curve shows the forecast CDF and the dashed orange line represents the Heaviside step function of the observation. The twCRPS is the integrated squared difference between the solid blue curve and the dashed orange line, with interval of integration ($x > t$) indicated by the green shaded region. \textbf{(b)} The threshold weight function $w(z) = \mathbb{1}(z > t)$. \textbf{(c)} The corresponding chaining function $v(z) = \max(z, t)$.}
    \label{fig:twcrps_graphical}
  \end{figure}

Figure~\ref{fig:twcrps_graphical} provides a graphical illustration of how the twCRPS is computed when a constant weight is applied to the upper tail of decision thresholds, aiding interpretation.

Recently, \citet{Allen_2023} showed that the twCRPS can be adapted for use with ensemble forecasts. The twCRPS for evaluating an ensemble forecast $F_{\mathrm{ens}}$ is defined as
\begin{equation}
    \mathrm{twCRPS}(F_{\mathrm{ens}}, y;v) = \frac{1}{M} \sum_{m=1}^M |v(x_m) - v(y)| - \frac{1}{2M^2} \sum_{m=1}^M \sum_{j=1}^M |v(x_m) - v(x_j)|,
    \label{eq:twCRPS-eNRG}
\end{equation}
\noindent where $v$ is the \emph{chaining function}. The chaining function $v$ is an antiderivative of the threshold weight function $w(z)$, such that
\begin{equation}
    v(z)-v(z') = \int_{z'}^{z} w(z) \,\mathrm{d}z.
\end{equation}

For example, if we wish to assign a threshold weight of 1 to thresholds above a specified threshold $t$, and a weight of 0 below $t$, the threshold weight function would be $w(z) = \mathbb{1}(z > t)$, where $\mathbb{1}$ is the indicator function that returns 1 if the condition is true and 0 otherwise. A corresponding chaining function is then $v(z) = \max(z, t)$. \citet{Allen_2023} and \citet{Allen_2024} provide further examples illustrating the relationship between chaining functions and threshold weight functions. 

Similarly to the fair CRPS, the fair twCRPS is defined as
\begin{equation}
    \text{twCRPS}_{\mathrm{fair}}(F_{ens}, y;v) = \frac{1}{M} \sum_{m=1}^M |v(x_m) - v(y)| - \frac{1}{2M(M-1)} \sum_{m=1}^M \sum_{j=1}^M |v(x_m) - v(x_j)|.
    \label{eq:twCRPS-fair}
\end{equation}

\subsection{Constructing and interpreting neighbourhood pseudo-ensembles}

When HiRA is used to compare forecast systems with different grid spacings, neighbourhoods can be selected to cover approximately the same physical area. For example, if Model A has a grid spacing of 3 km and Model B has a grid spacing of 9 km, a 27$\times$27 km neighbourhood can be represented by a 9$\times$9 grid of Model A and a 3$\times$3 grid of Model B, both centred on the observation point.

The score used to evaluate a neighbourhood pseudo-ensemble depends on how the neighbourhood forecast is interpreted. In this study, the $M$ native-grid values within a specified neighbourhood $\mathcal{N}$ define the equally weighted empirical CDF
\begin{equation}
\widehat F_{\mathcal{N}}(z)
=
\frac{1}{M}\sum_{m=1}^{M}\mathbb{1}{x_m\leq z}.
\label{eq:neighbourhood-ecdf}
\end{equation}
Under this interpretation, the empirical CDF is the forecast being evaluated. Its CRPS and twCRPS are therefore calculated using Eqs.~\ref{eq:CRPS-eNRG} and \ref{eq:twCRPS-eNRG}. This corresponds to using the values surrounding an observation site to construct a predictive distribution for that site.

Alternatively, the neighbourhood values can be treated as a finite sample from an underlying predictive distribution. The fair CRPS and twCRPS in Eqs.~\ref{eq:CRPS-fair} and \ref{eq:twCRPS-fair} account for the finite number of members under this interpretation, provided that the values can be treated as suitable draws from the underlying distribution. Because neighbouring grid values are spatially dependent, the suitability of this interpretation should be considered for each application. The fair correction does not remove differences in grid spacing or spatial scale.

Evaluating native-grid neighbourhoods compares the resulting forecast products, including effects associated with grid spacing, effective resolution, and spatial sampling. If the aim is instead to compare forecast systems at a common spatial scale, the forecasts should first be re-gridded to a common grid and evaluated against observations at a comparable spatial scale. In this study, we use the native-grid empirical-CDF interpretation and use the corresponding empirical scores. The Brier scores are also calculated from the same empirical forecast probabilities.

\section{CRPS and twCRPS HiRA results}
We select HRRR and GraphCast-GFS neighbourhoods that cover similar physical areas. Because GraphCast-GFS uses a latitude--longitude grid, its physical grid spacing varies with latitude, and no single HRRR neighbourhood covers exactly the same physical area at every station. We therefore use rectangular HRRR neighbourhoods that approximate the physical areas covered by the GraphCast-GFS neighbourhoods over CONUS. This differs from most HiRA applications, which typically use square or circular neighbourhoods. Table~\ref{tab:neighbourhoods} lists the paired neighbourhood sizes and their approximate physical dimensions.

Note that, although there is no neighbourhood size for GraphCast-GFS that is equivalent to the 3$\times$3 km point forecast for HRRR, we still evaluate the point-based HRRR to understand the impact of neighbourhood size on the scores because of the double penalty effect. The CRPS of a single ensemble member (i.e., the point-based forecast using a neighbourhood size of 1) is equivalent to the absolute loss. Likewise, threshold-weighted MAE (or threshold-weighted absolute loss) \citep{Taggart_2021} is equivalent to twCRPS for a single ensemble member.

When aggregating results spatially, we weight the results based on station density by following the approach in \citet{Rodwell_2010}. This ensures that the verification results are not overly influenced by geographical regions that have a higher concentration of stations compared to others. The station density weights are allowed to vary for each timestep to account for non-continuous reporting from weather stations.

\begin{table}[t]
\caption{Neighbourhood sizes used for HRRR and GraphCast-GFS, paired to cover similar physical areas.}
\begin{tabular}{lccc}
\tophline
Approximate physical dimensions (km) & HRRR grid points & GraphCast-GFS grid points\\
\middlehline
3$\times$3km & 1$\times$1 & - \\
21$\times$27km & 7$\times$9 & 1$\times$1  \\
63$\times$81km & 21$\times$27 & 3$\times$3 \\
\bottomhline
\end{tabular}
\label{tab:neighbourhoods}
\end{table}

\subsection{CRPS results}
We first calculate the mean CRPS of the specified neighbourhood-derived empirical forecasts to assess performance when all precipitation thresholds are weighted equally. Confidence intervals for the difference in mean scores between the two models are calculated as follows: spatial means are first taken across stations to account for spatial correlation, and the Hering–Genton modification of the Diebold–Mariano test \citep{Diebold_1995, Hering_2011} is then applied to account for temporal correlation.

\begin{figure}[htbp]
    \centering
    \includegraphics[width=0.75\textwidth]{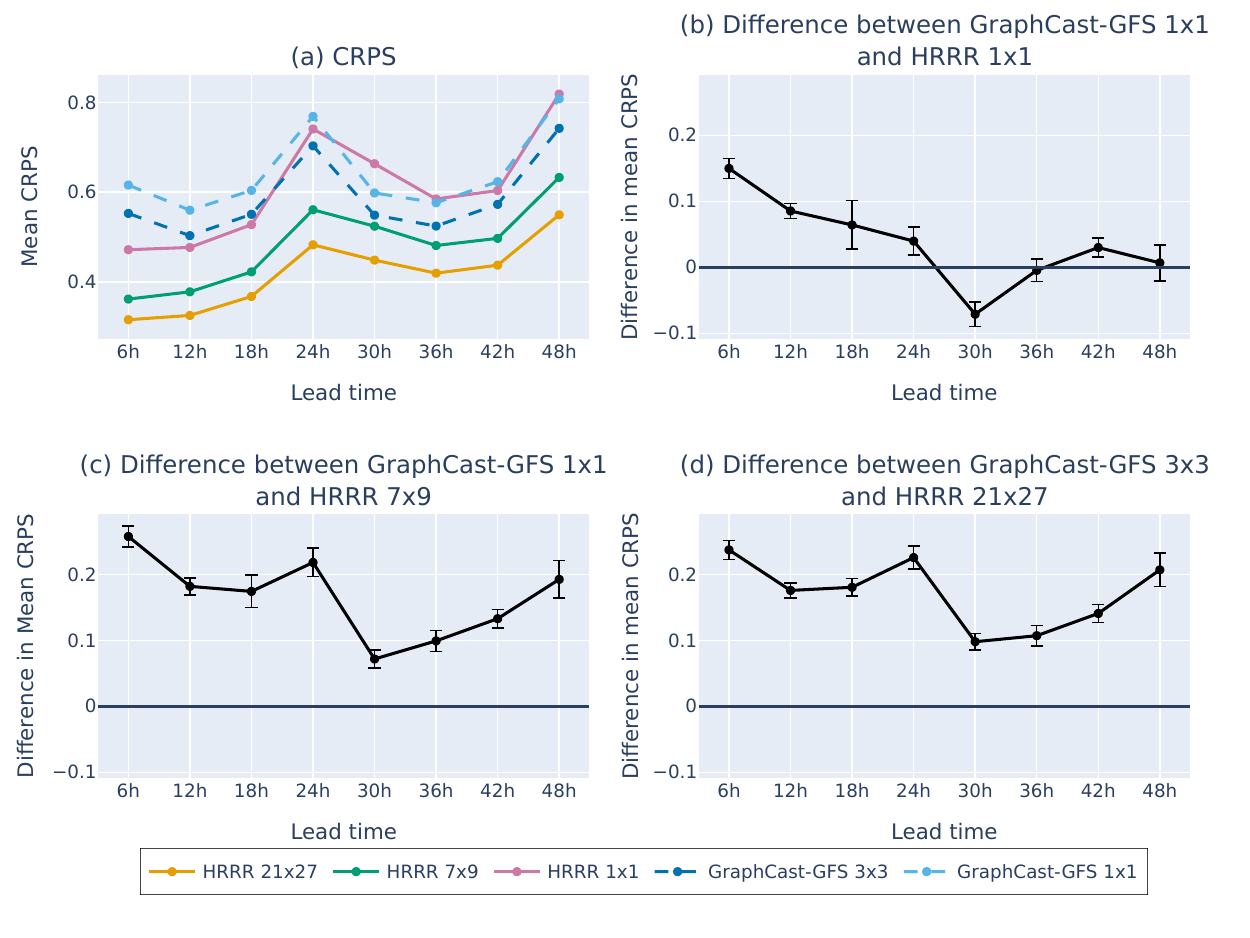}
    \caption{\textbf{(a)} Mean CRPS results aggregated across all stations and timesteps. Lower scores are better. \textbf{(b)} Difference between GraphCast-GFS 1$\times$1 and HRRR 1$\times$1 with 99\% confidence intervals. \textbf{(c)} Difference between GraphCast-GFS 1$\times$1 and HRRR 7$\times$9 ($21\times27$ km equivalent) with 99\% confidence intervals. \textbf{(d)} Difference between GraphCast-GFS 3$\times$3 and HRRR 21$\times$27 (63$\times$81 km equivalent) with 99\% confidence intervals. In subfigures b--d, positive values indicate that HRRR performed better than GraphCast-GFS for the specified neighbourhoods.} 
    \label{fig:crps_results}
  \end{figure}

Figure~\ref{fig:crps_results}a shows the mean CRPS for GraphCast-GFS and HRRR across various neighbourhood sizes. Increasing the neighbourhood size leads to a better score for both models. While one might expect a large improvement in performance when increasing the neighbourhood size in the HRRR as it is a high-resolution physical NWP model, an improvement also occurs with GraphCast-GFS despite its smoother appearance. A diurnal trend in model performance can also be seen. When the two models are compared gridpoint-to-gridpoint (Fig.~\ref{fig:crps_results}b), HRRR outperforms GraphCast-GFS during the first 24 hours, with mixed results at longer lead times. However, when the models are assessed over comparable neighbourhood sizes (Figs.~\ref{fig:crps_results}c-d), the HRRR demonstrated better performance across all lead times. 

\subsection{twCRPS results} 
To assess performance in predicting extremes, we construct a twCRPS weight function that takes the form $w(z) = \mathbb{1}(z>q_\alpha)$, where $q_\alpha$ is the $\alpha$ quantile of the climatological values in the ERA5 dataset for the point at which the ASOS station is located. We set $\alpha=0.99$ to focus on performance above the climatological 99th percentile. The corresponding chaining function that we use is $v(z) = \max(z, q_\alpha)$. When the mean twCRPS is calculated across time and space, $q_\alpha$ varies by station but is fixed in time. To test the impact of a more extreme threshold, we repeat the computation in Appendix A with $\alpha=0.999$ to examine performance above the climatological 99.9th percentile.

\begin{figure}[htbp]
    \centering
    \includegraphics[width=0.75\textwidth]{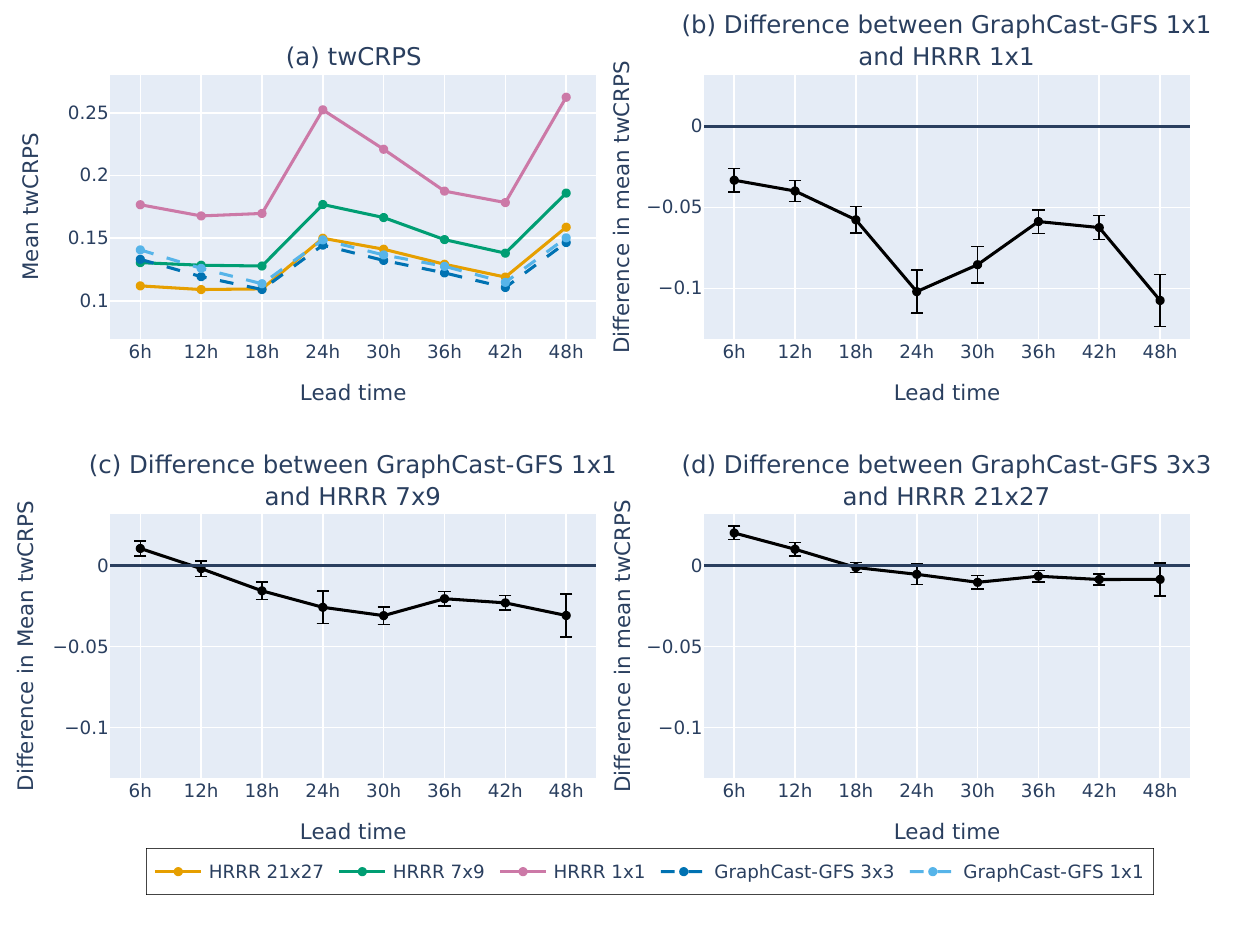}
    \caption{As for Fig.~\ref{fig:crps_results} but for the twCRPS with a threshold weight function of $w(z) = \mathbb{1}(z>q_{0.99})$.}
    \label{fig:twcrps_099_results}
  \end{figure}

In contrast to Fig.~\ref{fig:crps_results}a, Fig.~\ref{fig:twcrps_099_results}a shows that increasing the neighbourhood size had a larger impact on the mean twCRPS for the HRRR compared to GraphCast-GFS. When evaluated on a point-to-point basis (Fig.~\ref{fig:twcrps_099_results}b), GraphCast-GFS consistently achieved lower (better) twCRPS scores across all lead times. When evaluated across approximately equivalent neighbourhood sizes (Fig.~\ref{fig:twcrps_099_results}c-d), the HRRR performs better at shorter lead times, but not at longer lead times. Potentially, this may be attributed to the HRRR's assimilation of radar data, enhancing its short-term prediction of heavy precipitation. We leave a detailed investigation of this behaviour for future research.

\section{Decomposing CRPS across decision thresholds}
\begin{figure}[htbp]
    \centering
    \includegraphics[width=0.7\textwidth]{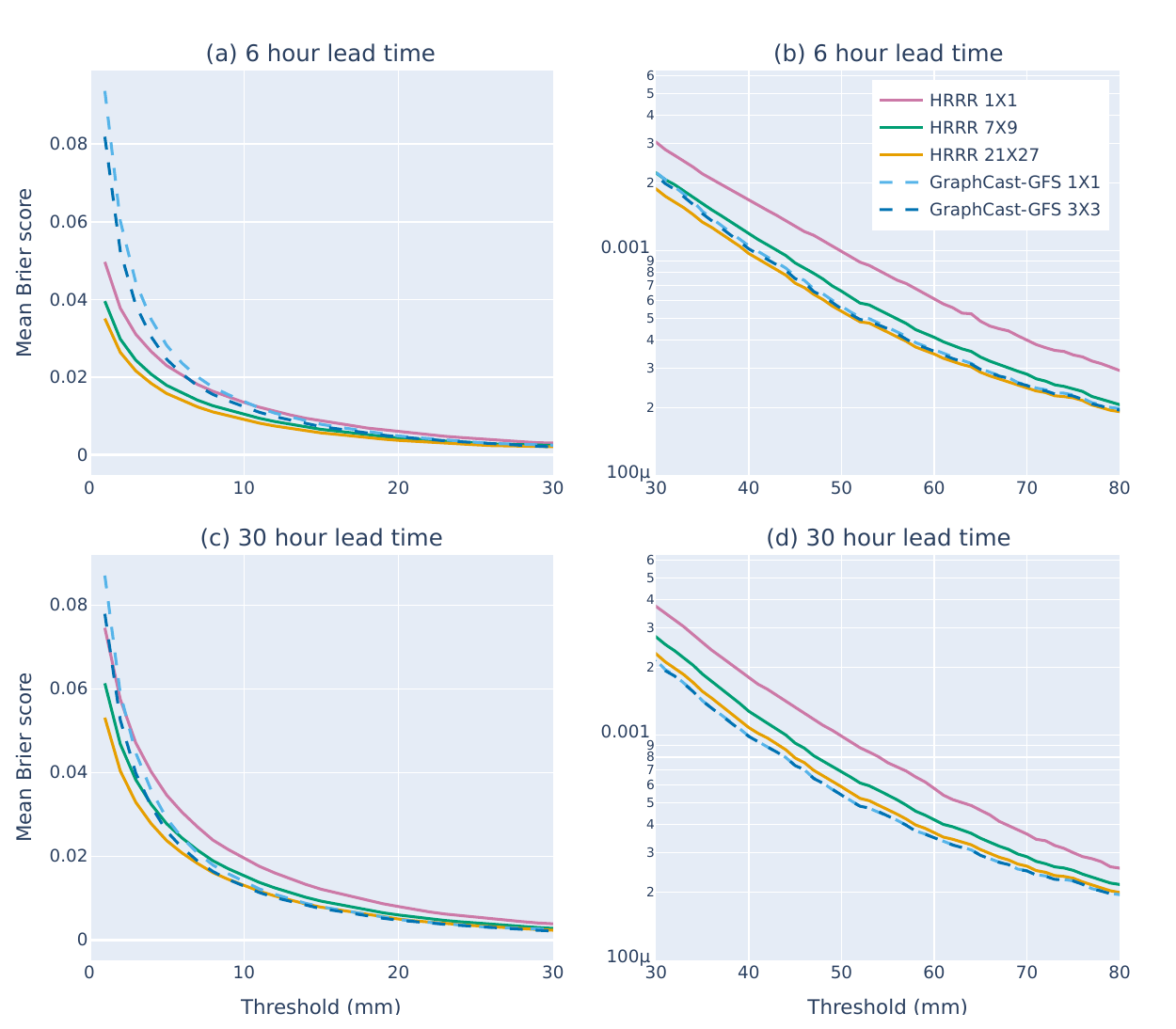}
    \caption{Brier score decomposition of the CRPS within the HiRA framework. Lower scores are better. The left panels \textbf{(a)} and \textbf{(c)} show the mean Brier score for thresholds below 30 mm, while the right panels \textbf{(b)} and \textbf{(d)} show the mean Brier score for threshold between 30 and 80~mm with a logarithmic vertical axis. Results are shown for six-hour lead time forecasts in panels \textbf{(a)} and \textbf{(b)}, and lead time 30 hour forecasts in \textbf{(c)} and \textbf{(d)}.}
    \label{fig:brier_results}
  \end{figure}

Since the CRPS is the integral of the Brier score across all thresholds (Eq.~\ref{eq:CRPS-Brier}), we decompose it to visualise the Brier scores across a range of thresholds. This provides us with greater insight within the HiRA framework as to how different models provide value at different decision thresholds. We display results for two lead times in Fig.~\ref{fig:brier_results}.

Figure~\ref{fig:brier_results} shows that the mean CRPS is dominated by non-extreme precipitation amounts (i.e., mean Brier score values from lower thresholds), which is unsurprising given their higher climatological frequency. To account for this, results are split at the 30~mm threshold, and the right-hand panels use a logarithmic vertical axis to better visualise the forecast rankings.

At a six-hour lead time (top panels), HRRR had lower mean Brier scores than GraphCast-GFS for precipitation thresholds below approximately 5~mm across all neighbourhood sizes. However, at the 30-hour lead time, this is only true for some neighbourhood sizes. In cases where the mean Brier score curves for HRRR and GraphCast-GFS intersect, the crossing point typically occurs at higher precipitation thresholds for larger HRRR neighbourhood sizes. This may reflect the increased likelihood of double-penalty effects at higher precipitation amounts when using point-to-point verification.
  
For the upper precipitation thresholds, HRRR 21$\times$27 had a slightly lower mean Brier score than GraphCast-GFS 3$\times$3 at the six-hour lead time and a slightly higher mean Brier score at the 30-hour lead time. This is consistent with the corresponding twCRPS results in Fig.~\ref{fig:twcrps_099_results}.

\section{Model climatology}
\begin{figure}[htbp]
    \centering
    \includegraphics[width=0.75\textwidth]{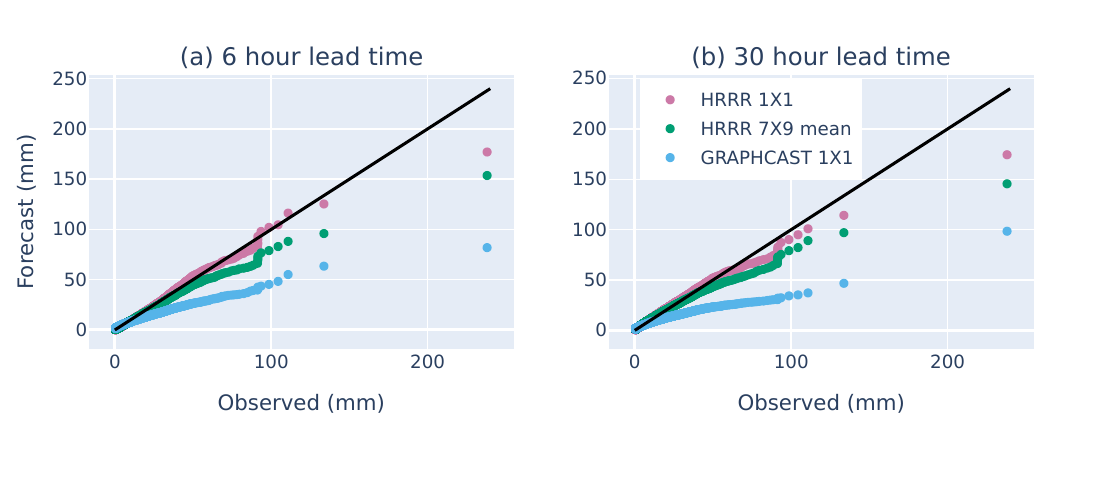}
    \caption{Q-Q plots of observations against forecasts. \textbf{(a)} shows results for six-hour lead time forecasts and \textbf{(b)} shows results for 30-hour lead times.}
    \label{fig:q-q}
  \end{figure}

We now assess the agreement between the model climatology and observations to understand any model biases. Quantile--Quantile (Q--Q) plots for the six-hour and 30-hour lead-time point-based forecasts and the observations are shown in Fig.~\ref{fig:q-q}. They show that the climatology of HRRR 1$\times$1 closely matches the observed climatology from the ASOS observations, whereas GraphCast-GFS 1$\times$1 is substantially less likely to predict heavier precipitation. This may partly reflect differences in grid resolution, with the HRRR model producing forecasts that correspond more closely to station-based observations, while GraphCast-GFS produces forecasts that may be more representative of a 0.25$^{\circ}$ grid. This is consistent with the issues involved in comparing gridded model precipitation forecasts against rain gauge observations highlighted by \citet{Tustison_2001}. To explore the relationship between grid resolution and representativeness, we also take the neighbourhood mean of HRRR 7$\times$9. The mean of the HRRR 7$\times$9 neighbourhood shifts the distribution away from the diagonal line, but not nearly to the same extent as GraphCast-GFS 1$\times$1, suggesting that differences in the climatology of model extremes are partly driven by the smoothness of GraphCast-GFS and not just because of differences in grid resolution.

\section{Discrimination ability}
\begin{figure}[htbp]
    \centering
    \includegraphics[width=0.7\textwidth]{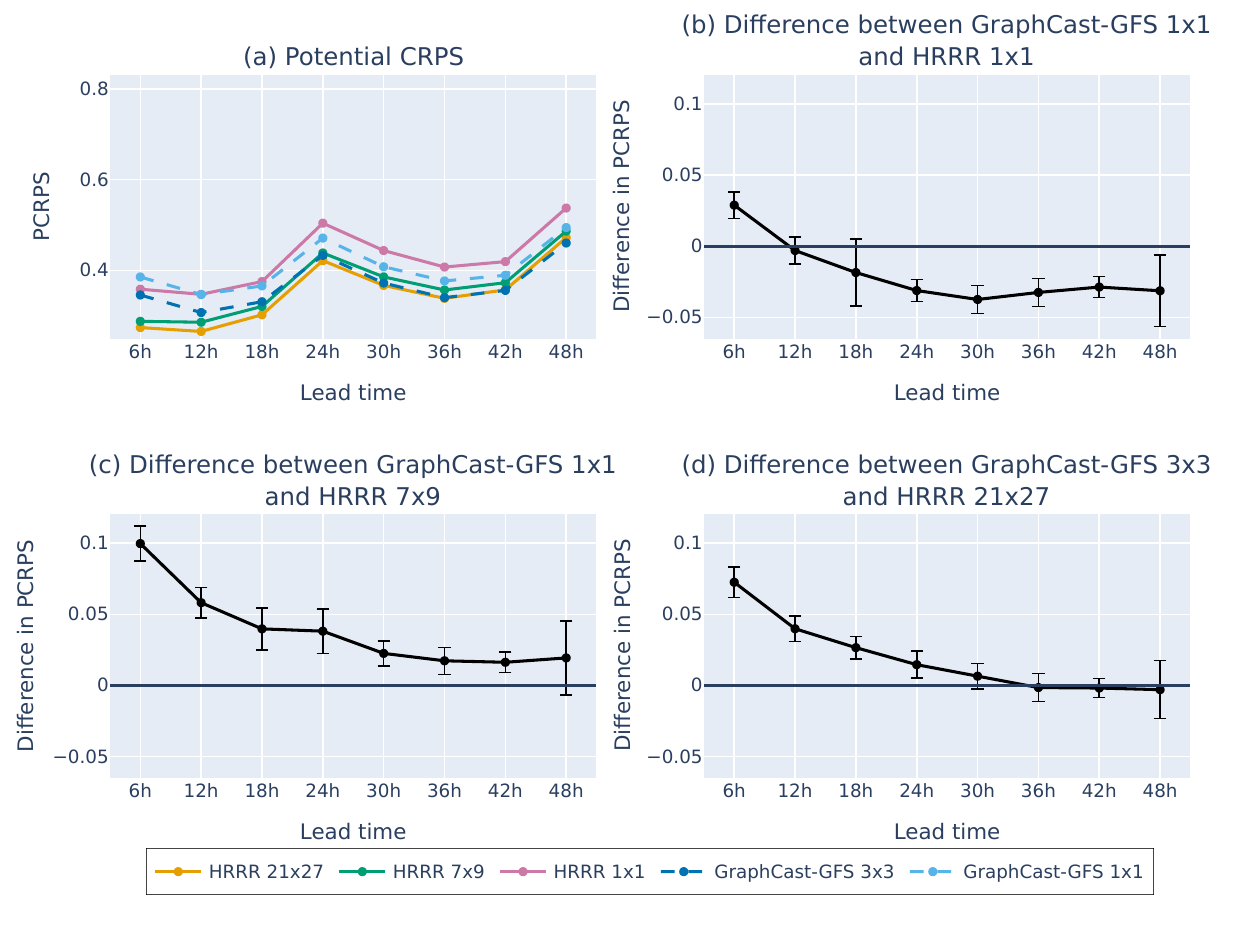}
    \caption{\textbf{(a)} Potential CRPS results. Lower values indicate more discrimination ability when all decision thresholds are weighted equally. \textbf{(b)} Difference between GraphCast-GFS 1$\times$1 and HRRR 1$\times$1 with 99\% confidence intervals. \textbf{(c)} Difference between GraphCast-GFS 1$\times$1 and HRRR 7$\times$9 with 99\% confidence intervals. \textbf{(d)} Difference between GraphCast-GFS 3$\times$3 and HRRR 21$\times$27 with 99\% confidence intervals. In subfigures b--d, positive values indicate that HRRR had more discrimination ability than GraphCast-GFS for the specified neighbourhoods.}
    \label{fig:potential_crps}
  \end{figure}

\begin{figure}[htbp]
    \centering
    \includegraphics[width=0.7\textwidth]{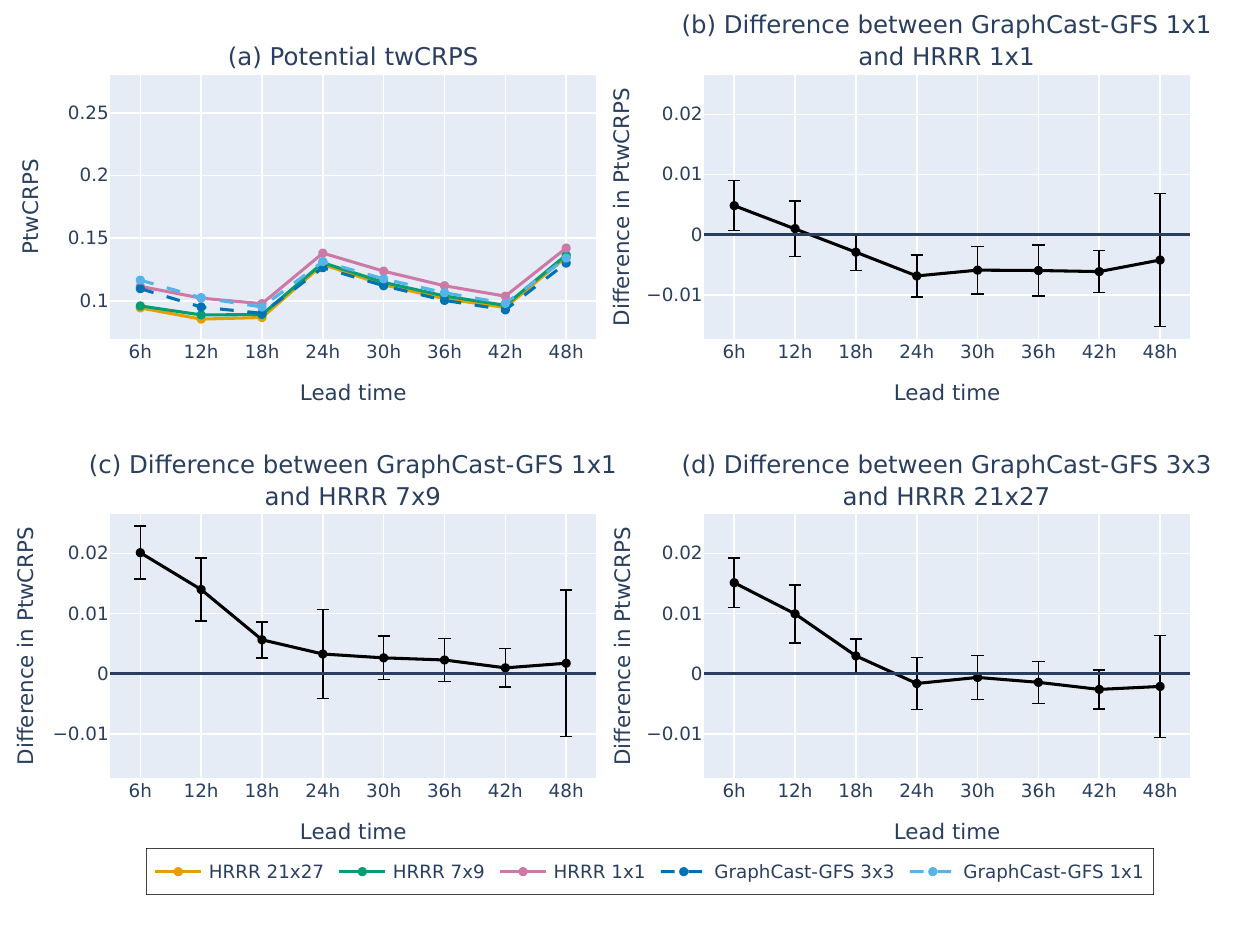}
    \caption{\textbf{(a)} As for Fig.~\ref{fig:potential_crps}, but for potential twCRPS. Lower values indicate more discrimination ability for predicting extremes (all thresholds above the climatological 99th percentile).}
    \label{fig:potential_twcrps}
  \end{figure}

As demonstrated in the preceding section, the behaviour of GraphCast-GFS may not be consistent with in-situ observations. For this reason, it is important to understand the discrimination ability (i.e., potential predictive ability) of the models as a way of accounting for the models representing the precipitation across a grid cell rather than at a specific point. This is because meteorologists may be able to learn how the model behaves and use its output accordingly. Alternatively, a model with good discrimination ability could easily be post-processed to correct for any conditional biases. 
The concept of ``potential'' scores as a measure of potential predictive or discrimination ability has been proposed and extended multiple times, including for the CRPS \citep[e.g.,][]{Murphy_1989, Hersbach_2000, Br_cker_2011, Siegert_2017}. Recently \citet{gneiting2025probabilisticmeasuresaffordfair} introduced a version of a potential CRPS to measure and compare the discrimination ability of a deterministic NWP model to that of an AIWP model. This was done by calibrating the model (in-sample) using IDR via the EasyUQ method \citep{Walz_2024} to convert the single-valued forecasts to probabilistic forecasts. We adopt a related, but distinct approach for calculating the ``potential'' CRPS and twCRPS. Instead of using IDR on each neighbourhood member, we apply isotonic regression (in-sample) to each neighbourhood member. Isotonic regression is a method for fitting a nondecreasing function to a set of forecast-observation pairs \citep{Ayer_1955}. Specifically, we use isotonic median regression (isotonic regression under L1 loss), which targets the median functional.
We adopt this approach to approximate the behaviour of an operational meteorologist who is familiar with typical model biases and applies simple calibrations to the model output. Additionally, it partly circumvents the issue in neighbourhood verification approaches where adjacent grid cells in the neighbourhood are not representative of the station site (e.g., in areas of varying topography). This approach, however, is unlikely to be as effective as IDR in producing a well-calibrated predictive distribution if neighbourhood members were used as covariates. We then calculate a potential CRPS and potential twCRPS using the forecasts calibrated via isotonic regression as shown in Fig.~\ref{fig:potential_crps} and Fig.~\ref{fig:potential_twcrps}.

The potential CRPS and potential twCRPS results, which measure the discrimination ability, show that while increasing neighbourhood size leads to better discrimination ability, the HRRR and GraphCast-GFS are sometimes ranked differently compared to Fig.~\ref{fig:crps_results} and Fig.~\ref{fig:twcrps_099_results}, which measure overall predictive performance rather than discrimination ability.

The potential CRPS results indicate that, when all decision thresholds are weighted equally, 
the HRRR $7\times9$ shows stronger discrimination ability than GraphCast-GFS $1\times1$ 
across all lead times, and the HRRR $21\times27$ exhibits stronger discrimination ability 
than GraphCast-GFS $3\times3$ at short lead times, with the discrimination ability of the two models converging thereafter.

Examining discrimination ability for extreme decision thresholds based on the potential twCRPS results, the HRRR model outperforms 
GraphCast-GFS at a lead time of 6 hours for all equivalent neighbourhood sizes; however, this 
advantage diminishes with increasing lead time, and GraphCast-GFS exhibits greater 
discrimination ability at lead times of 24 hours and beyond for the GraphCast-GFS $3\times3$ versus HRRR $21\times27$ comparison. Differences in discrimination ability for the 
latter comparison were not statistically significant.

\conclusions  
This paper demonstrates an approach to evaluating how an AIWP model compares to a high-resolution physical NWP model in predicting climatologically extreme six-hourly precipitation. It combines two existing techniques: the HiRA framework and twCRPS. Model performance was assessed within a framework representative of potential operational use by meteorologists or simple post-processing systems. As with all NWP verification methods, this approach does not definitively determine which model is superior across all use cases. The comparison includes, rather than removes, differences associated with grid spacing, effective resolution, and spatial sampling. It is not a comparison after re-gridding the forecast systems to a common spatial scale.

When models were compared using neighbourhood sizes that covered a similar physical area to generate empirical CDFs, HRRR consistently outperformed GraphCast-GFS across all lead times as measured by CRPS. However, when focusing on predictive performance of extreme precipitation, HRRR only outperformed GraphCast-GFS at short lead times.

Decomposing CRPS by decision threshold provided further insight into which thresholds each model handled more effectively. Where Brier score curves for HRRR and GraphCast-GFS intersected, this generally occurred at higher precipitation thresholds for larger HRRR neighbourhood sizes.

The approach was extended to measure the discrimination ability of the models to predict extreme precipitation. This was important to measure since GraphCast-GFS forecasts of heavy precipitation showed an under-forecast bias when evaluated against rain gauge data. Increasing the neighbourhood size led to superior discrimination ability. When GraphCast-GFS 3$\times$3 versus HRRR 21$\times$27 were compared, the high-resolution model had superior discrimination ability at short lead times, but not beyond 24 hours.

This verification approach has several strengths:
\begin{enumerate}
    \item It is a spatial method that can address the double penalty effect within a specified distance.
    \item It allows forecast systems with different grid spacings to be evaluated on their native grids when the forecast being assessed is defined as the finite neighbourhood empirical distribution.
    \item It can be used to evaluate models against in-situ observations.
    \item It can be threshold-weighted to focus on extremes or other important decision thresholds.
    \item It uses proper scoring rules within a user-focused framework, avoiding selection bias associated with conditioning on extreme observed or forecast events.
    \item It supports aggregation across domains by accounting for climatological differences via threshold weighting if required.
\end{enumerate}

As noted by \citet{pic2024properscoringrulesmultivariate}, there is an opportunity to bridge the gap between the spatial verification methods and proper scoring rules communities. This work offers one such integration, and raises several opportunities for future research:
\begin{itemize}
    \item The approach could be extended to evaluate multivariate forecasts by using threshold-weighted multivariate scores, such as the threshold-weighted variogram score \citep{Allen_2023} which is a proper scoring rule. This approach could be used to evaluate compound events as well as different variables simultaneously (e.g., surface wind speed, temperature and relative humidity for fire weather).
    \item Comparing this approach to other spatial verification methods.
    \item Applying threshold-weighted scoring rules within HiRA to ensemble forecasts, as in \citet{Mittermaier_2017}.
    \item Extending the approach to also account for timing errors.
    \item Applying neighbourhood twCRPS to forecasts evaluated with gridded rainfall observations, which is similar to \citet{Stein_2022} who used a neighbourhood CRPS with gridded data. 
    \item Comparing a potential CRPS measure \citep{gneiting2025probabilisticmeasuresaffordfair} that applies IDR using neighbourhood members as covariates. This may be useful when models are spatially sharper.
\end{itemize}

Finally, while most prior comparisons between AIWP and traditional NWP models have relied on point-to-point verification, this study shows that model rankings can differ when assessed from a spatial perspective, consistent with findings from \citet{Radford_2025_comparison}. If AIWP models are to be operationalised, adopting spatial verification methods aligned with their practical use will be essential for accurately understanding their performance.


\codedataavailability{} 
All code to reproduce the analysis and figures in this paper is available at https://doi.org/10.5281/zenodo.20438050 \citep{loveday_2026_code}.

All data used to reproduce this work is available at https://zenodo.org/records/19672887 \citep{loveday_2026_data}.

The original data sources for the data stored in the above record were retrieved from the following sources:
\begin{itemize}
    \item One minute ASOS data were retrieved from https://mesonet.agron.iastate.edu/request/asos/1min.phtml which contains an archive of data provided by the National Climatic Data Center.
    \item HRRR data was retrieved from https://hrrrzarr.s3.amazonaws.com/index.html.
    \item GraphCast-GFS data was retrieved from https://noaa-oar-mlwp-data.s3.amazonaws.com/index.html.
    \item ERA5 data was retrieved from https://console.cloud.google.com/storage/browser/weatherbench2/data/era5.
\end{itemize}

The twCRPS, CRPS for ensembles, Brier score for ensembles, and Q-Q plots were added to the \textit{scores} Python package \citep{Leeuwenburg_2024} as part of this work and can also be found at https://github.com/nci/scores. The specific version of \textit{scores} used in this work was 2.5.0 \citep{leeuwenburg_2026_18638494}. Other methods including the Diebold Mariano (with the Hering-Genton modification) test and isotonic regression can also be found in the \textit{scores} Python package.

\section{Appendix A: twCRPS results for the 0.999 climatological threshold}
\begin{figure}[htbp]
    \centering
    \includegraphics[width=0.75\textwidth]{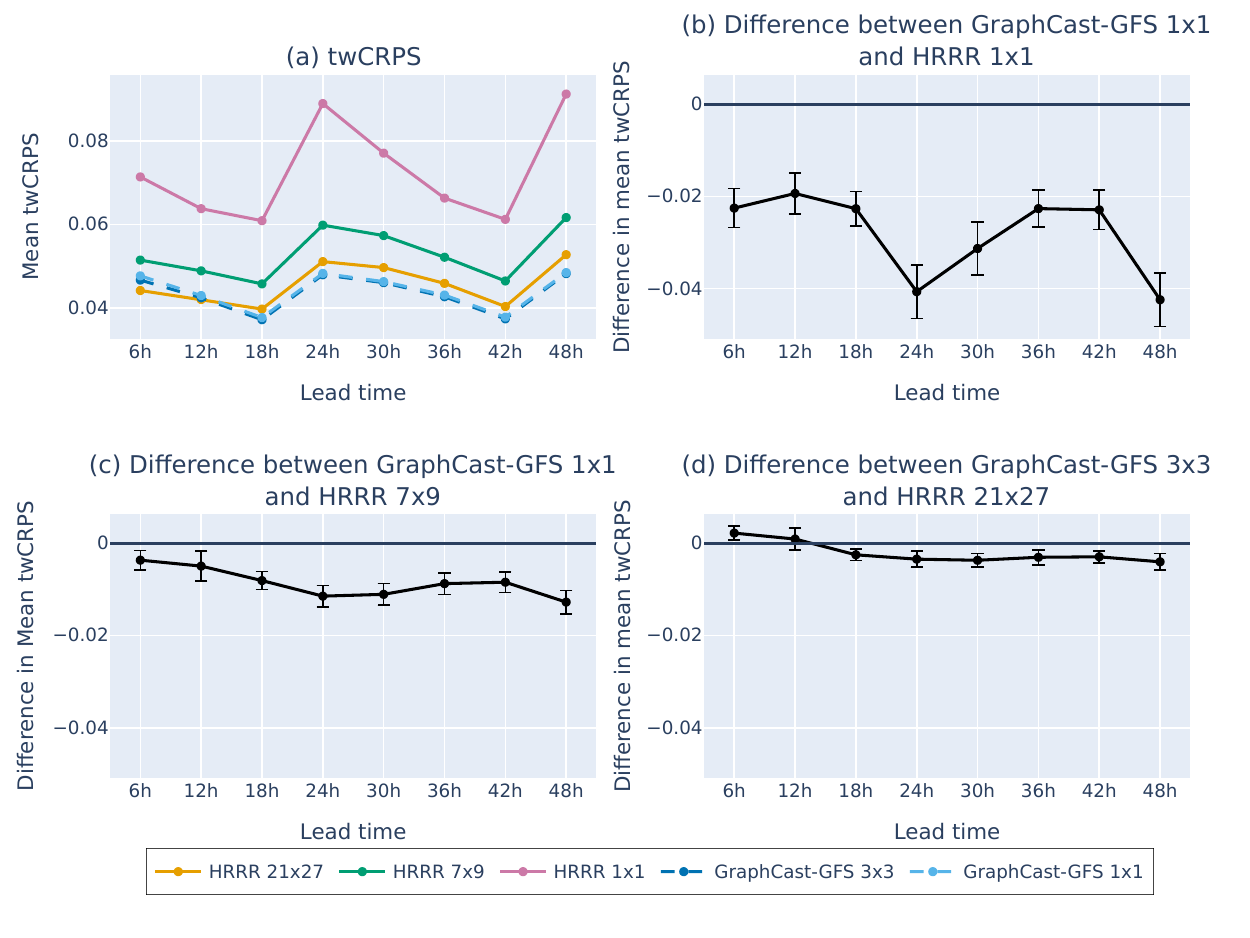}
    \caption{As for Fig.~\ref{fig:crps_results} but for the twCRPS with a threshold weight function of $w(z) = \mathbb{1}(z>q_{0.999})$.}
    \label{fig:twcrps_0999_results}
  \end{figure}

Results for the twCRPS using a 0.999 climatological threshold are shown in Fig.~\ref{fig:twcrps_0999_results}. The only instance in which the HRRR was significantly better than GraphCast-GFS was the six-hour lead time when GraphCast-GFS 3$\times$3 and HRRR 21$\times$27 were compared. Results were also calculated using a fixed 50~mm threshold at all locations and were similar (not shown).
\appendixfigures
  \clearpage
\noappendix       







\authorcontribution{\textbf{NL} Conceptualisation, Data Curation, Formal Analysis, Investigation, Methodology, Software, Visualisation, Writing - Original Draft Preparation, Writing - Review and Editing. \textbf{TH} Writing - Review and Editing, Project Administration, Resources.} 

\competinginterests{The authors have no conflict of interest to declare.} 

\begin{acknowledgements}
This research was funded by NSF NCAR's Developmental Testbed Center (DTC) Visiting Scientist Program. The authors thank Louisa Nance (NSF NCAR and DTC), Belinda Trotta (BoM), Robert Taggart (BoM), and two anonymous reviewers for feedback on earlier versions of this manuscript. Nicholas Loveday would like to thank everyone at NCAR and NOAA GSL who supported his visit as a DTC Visiting Scientist. 
\end{acknowledgements}







\bibliographystyle{copernicus}
\bibliography{example.bib}

@misc{abdi2025hrrrcastdatadrivenemulatorregional,
  title         = {HRRRCast: a data-driven emulator for regional weather forecasting at convection allowing scales},
  author        = {Daniel Abdi and Isidora Jankov and Paul Madden and Vanderlei Vargas and Timothy A. Smith and Sergey Frolov and Montgomery Flora and Corey Potvin},
  year          = {2025},
  eprint        = {2507.05658},
  archiveprefix = {arXiv},
  primaryclass  = {physics.ao-ph},
  url           = {https://arxiv.org/abs/2507.05658}
}

@misc{adamov2025buildingmachinelearninglimited,
  title         = {Building Machine Learning Limited Area Models: Kilometer-Scale Weather Forecasting in Realistic Settings},
  author        = {Simon Adamov and Joel Oskarsson and Leif Denby and Tomas Landelius and Kasper Hintz and Simon Christiansen and Irene Schicker and Carlos Osuna and Fredrik Lindsten and Oliver Fuhrer and Sebastian Schemm},
  year          = {2025},
  eprint        = {2504.09340},
  archiveprefix = {arXiv},
  primaryclass  = {physics.ao-ph},
  url           = {https://arxiv.org/abs/2504.09340}
}

@article{Allen_2023,
  title     = {Evaluating Forecasts for High-Impact Events Using Transformed Kernel Scores},
  volume    = {11},
  issn      = {2166-2525},
  url       = {http://dx.doi.org/10.1137/22m1532184},
  doi       = {10.1137/22m1532184},
  number    = {3},
  journal   = {SIAM/ASA Journal on Uncertainty Quantification},
  publisher = {Society for Industrial & Applied Mathematics (SIAM)},
  author    = {Allen, Sam and Ginsbourger, David and Ziegel, Johanna},
  year      = {2023},
  month     = aug,
  pages     = {906–940}
}

@article{Allen_2024,
  title     = {Weighted scoringRules: Emphasizing Particular Outcomes When Evaluating Probabilistic Forecasts},
  volume    = {110},
  issn      = {1548-7660},
  url       = {http://dx.doi.org/10.18637/jss.v110.i08},
  doi       = {10.18637/jss.v110.i08},
  number    = {8},
  journal   = {Journal of Statistical Software},
  publisher = {Foundation for Open Access Statistic},
  author    = {Allen, Sam},
  year      = {2024}
}

@article{Ayer_1955,
  title     = {An Empirical Distribution Function for Sampling with Incomplete Information},
  volume    = {26},
  issn      = {0003-4851},
  url       = {http://dx.doi.org/10.1214/aoms/1177728423},
  doi       = {10.1214/aoms/1177728423},
  number    = {4},
  journal   = {The Annals of Mathematical Statistics},
  publisher = {Institute of Mathematical Statistics},
  author    = {Ayer, Miriam and Brunk, H. D. and Ewing, G. M. and Reid, W. T. and Silverman, Edward},
  year      = {1955},
  month     = dec,
  pages     = {641–647}
}

@article{Baringhaus_2004,
  title     = {On a new multivariate two-sample test},
  volume    = {88},
  issn      = {0047-259X},
  url       = {http://dx.doi.org/10.1016/s0047-259x(03)00079-4},
  doi       = {10.1016/s0047-259x(03)00079-4},
  number    = {1},
  journal   = {Journal of Multivariate Analysis},
  publisher = {Elsevier BV},
  author    = {Baringhaus, L. and Franz, C.},
  year      = {2004},
  month     = jan,
  pages     = {190–206}
}

@article{Ben_Bouall_gue_2024,
  title     = {The Rise of Data-Driven Weather Forecasting: A First Statistical Assessment of Machine Learning–Based Weather Forecasts in an Operational-Like Context},
  volume    = {105},
  issn      = {1520-0477},
  url       = {http://dx.doi.org/10.1175/bams-d-23-0162.1},
  doi       = {10.1175/bams-d-23-0162.1},
  number    = {6},
  journal   = {Bulletin of the American Meteorological Society},
  publisher = {American Meteorological Society},
  author    = {Ben Bouallègue, Zied and Clare, Mariana C. A. and Magnusson, Linus and Gascón, Estibaliz and Maier-Gerber, Michael and Janoušek, Martin and Rodwell, Mark and Pinault, Florian and Dramsch, Jesper S. and Lang, Simon T. K. and Raoult, Baudouin and Rabier, Florence and Chevallier, Matthieu and Sandu, Irina and Dueben, Peter and Chantry, Matthew and Pappenberger, Florian},
  year      = {2024},
  month     = jun,
  pages     = {E864–E883}
}

@article{Ben_Bouall_gue_2026, title={SEEPS4ALL: an open dataset for the verification of daily precipitation forecasts using station climate statistics}, volume={18}, ISSN={1866-3516}, url={http://dx.doi.org/10.5194/essd-18-713-2026}, DOI={10.5194/essd-18-713-2026}, number={1}, journal={Earth System Science Data}, publisher={Copernicus GmbH}, author={Ben-Bouallègue, Zied and Prieto-Nemesio, Ana and Iza Wong, Angela and Pinault, Florian and van der Schee, Marlies and Modigliani, Umberto}, year={2026}, month=Jan, pages={713–720} }

@article{Bi_2023,
  title     = {Accurate medium-range global weather forecasting with 3D neural networks},
  volume    = {619},
  issn      = {1476-4687},
  url       = {http://dx.doi.org/10.1038/s41586-023-06185-3},
  doi       = {10.1038/s41586-023-06185-3},
  number    = {7970},
  journal   = {Nature},
  publisher = {Springer Science and Business Media LLC},
  author    = {Bi, Kaifeng and Xie, Lingxi and Zhang, Hengheng and Chen, Xin and Gu, Xiaotao and Tian, Qi},
  year      = {2023},
  month     = jul,
  pages     = {533–538}
}

@article{Brenowitz_2025,
  title     = {A Practical Probabilistic Benchmark for AI Weather Models},
  volume    = {52},
  issn      = {1944-8007},
  url       = {http://dx.doi.org/10.1029/2024gl113656},
  doi       = {10.1029/2024gl113656},
  number    = {7},
  journal   = {Geophysical Research Letters},
  publisher = {American Geophysical Union (AGU)},
  author    = {Brenowitz, Noah D. and Cohen, Yair and Pathak, Jaideep and Mahesh, Ankur and Bonev, Boris and Kurth, Thorsten and Durran, Dale R. and Harrington, Peter and Pritchard, Michael S.},
  year      = {2025},
  month     = apr
}

@article{BRIER_1950,
  title     = {VERIFICATION OF FORECASTS EXPRESSED IN TERMS OF PROBABILITY},
  volume    = {78},
  issn      = {1520-0493},
  url       = {http://dx.doi.org/10.1175/1520-0493(1950)078<0001:vofeit>2.0.co;2},
  doi       = {10.1175/1520-0493(1950)078<0001:vofeit>2.0.co;2},
  number    = {1},
  journal   = {Monthly Weather Review},
  publisher = {American Meteorological Society},
  author    = {Brier, Glenn W.},
  year      = {1950},
  month     = jan,
  pages     = {1–3}
}

@article{Br_cker_2011, title={Estimating reliability and resolution of probability forecasts through decomposition of the empirical score}, volume={39}, ISSN={1432-0894}, url={http://dx.doi.org/10.1007/s00382-011-1191-1}, DOI={10.1007/s00382-011-1191-1}, number={3-4}, journal={Climate Dynamics}, publisher={Springer Science and Business Media LLC}, author={Bröcker, Jochen}, year={2011}, month=Oct, pages={655–667} }

@article{Charlton_Perez_2024,
  title     = {Do AI models produce better weather forecasts than physics-based models? A quantitative evaluation case study of Storm Ciarán},
  volume    = {7},
  issn      = {2397-3722},
  url       = {http://dx.doi.org/10.1038/s41612-024-00638-w},
  doi       = {10.1038/s41612-024-00638-w},
  number    = {1},
  journal   = {npj Climate and Atmospheric Science},
  publisher = {Springer Science and Business Media LLC},
  author    = {Charlton-Perez, Andrew J. and Dacre, Helen F. and Driscoll, Simon and Gray, Suzanne L. and Harvey, Ben and Harvey, Natalie J. and Hunt, Kieran M. R. and Lee, Robert W. and Swaminathan, Ranjini and Vandaele, Remy and Volonté, Ambrogio},
  year      = {2024},
  month     = apr
}

@article{Diebold_1995, title={Comparing Predictive Accuracy}, volume={13}, ISSN={1537-2707}, url={http://dx.doi.org/10.1080/07350015.1995.10524599}, DOI={10.1080/07350015.1995.10524599}, number={3}, journal={Journal of Business \& Economic Statistics}, publisher={Informa UK Limited}, author={Diebold, Francis X. and Mariano, Roberto S.}, year={1995}, month=jul, pages={253–263} }

@article{Dorninger_2018,
  title     = {The Setup of the MesoVICT Project},
  volume    = {99},
  issn      = {1520-0477},
  url       = {http://dx.doi.org/10.1175/bams-d-17-0164.1},
  doi       = {10.1175/bams-d-17-0164.1},
  number    = {9},
  journal   = {Bulletin of the American Meteorological Society},
  publisher = {American Meteorological Society},
  author    = {Dorninger, Manfred and Gilleland, Eric and Casati, Barbara and Mittermaier, Marion P. and Ebert, Elizabeth E. and Brown, Barbara G. and Wilson, Laurence J.},
  year      = {2018},
  month     = sep,
  pages     = {1887–1906}
}

@article{Ebert_2009,
  title     = {Neighborhood Verification: A Strategy for Rewarding Close Forecasts},
  volume    = {24},
  issn      = {0882-8156},
  url       = {http://dx.doi.org/10.1175/2009waf2222251.1},
  doi       = {10.1175/2009waf2222251.1},
  number    = {6},
  journal   = {Weather and Forecasting},
  publisher = {American Meteorological Society},
  author    = {Ebert, Elizabeth E.},
  year      = {2009},
  month     = dec,
  pages     = {1498–1510}
}

@article{Dowell_2022,
  title     = {The High-Resolution Rapid Refresh (HRRR): An Hourly Updating Convection-Allowing Forecast Model. Part I: Motivation and System Description},
  volume    = {37},
  issn      = {1520-0434},
  url       = {http://dx.doi.org/10.1175/waf-d-21-0151.1},
  doi       = {10.1175/waf-d-21-0151.1},
  number    = {8},
  journal   = {Weather and Forecasting},
  publisher = {American Meteorological Society},
  author    = {Dowell, David C. and Alexander, Curtis R. and James, Eric P. and Weygandt, Stephen S. and Benjamin, Stanley G. and Manikin, Geoffrey S. and Blake, Benjamin T. and Brown, John M. and Olson, Joseph B. and Hu, Ming and Smirnova, Tatiana G. and Ladwig, Terra and Kenyon, Jaymes S. and Ahmadov, Ravan and Turner, David D. and Duda, Jeffrey D. and Alcott, Trevor I.},
  year      = {2022},
  month     = aug,
  pages     = {1371–1395}
}

@article{Ebert_2008,
  title     = {Fuzzy verification of high‐resolution gridded forecasts: a review and proposed framework},
  volume    = {15},
  issn      = {1469-8080},
  url       = {http://dx.doi.org/10.1002/met.25},
  doi       = {10.1002/met.25},
  number    = {1},
  journal   = {Meteorological Applications},
  publisher = {Wiley},
  author    = {Ebert, Elizabeth E.},
  year      = {2008},
  month     = mar,
  pages     = {51–64}
}

@article{Epstein_1969,
  title     = {A Scoring System for Probability Forecasts of Ranked Categories},
  volume    = {8},
  issn      = {0021-8952},
  url       = {http://dx.doi.org/10.1175/1520-0450(1969)008<0985:assfpf>2.0.co;2},
  doi       = {10.1175/1520-0450(1969)008<0985:assfpf>2.0.co;2},
  number    = {6},
  journal   = {Journal of Applied Meteorology},
  publisher = {American Meteorological Society},
  author    = {Epstein, Edward S.},
  year      = {1969},
  month     = dec,
  pages     = {985–987}
}

@article{Ferro_2013,
  title     = {Fair scores for ensemble forecasts: Fair Scores for Ensemble Forecasts},
  volume    = {140},
  issn      = {0035-9009},
  url       = {http://dx.doi.org/10.1002/qj.2270},
  doi       = {10.1002/qj.2270},
  number    = {683},
  journal   = {Quarterly Journal of the Royal Meteorological Society},
  publisher = {Wiley},
  author    = {Ferro, C. A. T.},
  year      = {2013},
  month     = dec,
  pages     = {1917–1923}
}

@article{Fovell_2022,
  title     = {An Evaluation of Surface Wind and Gust Forecasts from the High-Resolution Rapid Refresh Model},
  volume    = {37},
  issn      = {1520-0434},
  url       = {http://dx.doi.org/10.1175/waf-d-21-0176.1},
  doi       = {10.1175/waf-d-21-0176.1},
  number    = {6},
  journal   = {Weather and Forecasting},
  publisher = {American Meteorological Society},
  author    = {Fovell, Robert G. and Gallagher, Alex},
  year      = {2022},
  month     = jun,
  pages     = {1049–1068}
}

@article{Gilleland_2009,
  title     = {Intercomparison of Spatial Forecast Verification Methods},
  volume    = {24},
  issn      = {0882-8156},
  url       = {http://dx.doi.org/10.1175/2009waf2222269.1},
  doi       = {10.1175/2009waf2222269.1},
  number    = {5},
  journal   = {Weather and Forecasting},
  publisher = {American Meteorological Society},
  author    = {Gilleland, Eric and Ahijevych, David and Brown, Barbara G. and Casati, Barbara and Ebert, Elizabeth E.},
  year      = {2009},
  month     = oct,
  pages     = {1416–1430}
}

@article{Gneiting_2007,
  title     = {Strictly Proper Scoring Rules, Prediction, and Estimation},
  volume    = {102},
  issn      = {1537-274X},
  url       = {http://dx.doi.org/10.1198/016214506000001437},
  doi       = {10.1198/016214506000001437},
  number    = {477},
  journal   = {Journal of the American Statistical Association},
  publisher = {Informa UK Limited},
  author    = {Gneiting, Tilmann and Raftery, Adrian E},
  year      = {2007},
  month     = mar,
  pages     = {359–378}
}

@article{Gneiting_2011,
  title     = {Comparing Density Forecasts Using Threshold- and Quantile-Weighted Scoring Rules},
  volume    = {29},
  issn      = {1537-2707},
  url       = {http://dx.doi.org/10.1198/jbes.2010.08110},
  doi       = {10.1198/jbes.2010.08110},
  number    = {3},
  journal   = {Journal of Business \& Economic Statistics},
  publisher = {Informa UK Limited},
  author    = {Gneiting, Tilmann and Ranjan, Roopesh},
  year      = {2011},
  month     = jul,
  pages     = {411–422}
}

@article{Gneiting_2011_point,
  title     = {Making and Evaluating Point Forecasts},
  volume    = {106},
  issn      = {1537-274X},
  url       = {http://dx.doi.org/10.1198/jasa.2011.r10138},
  doi       = {10.1198/jasa.2011.r10138},
  number    = {494},
  journal   = {Journal of the American Statistical Association},
  publisher = {Informa UK Limited},
  author    = {Gneiting, Tilmann},
  year      = {2011},
  month     = jun,
  pages     = {746–762}
}

@article{gneiting2014,
  author    = {Gneiting, Tilmann and Katzfuss, Matthias},
  title     = {Probabilistic Forecasting},
  journal   = {Annual Review of Statistics and Its Application},
  year      = {2014},
  volume    = {1},
  number    = {Volume 1, 2014},
  pages     = {125-151},
  doi       = {https://doi.org/10.1146/annurev-statistics-062713-085831},
  url       = {https://www.annualreviews.org/content/journals/10.1146/annurev-statistics-062713-085831},
  publisher = {Annual Reviews},
  issn      = {2326-831X},
  type      = {Journal Article}
}

@misc{gneiting2025probabilisticmeasuresaffordfair,
  title         = {Probabilistic measures afford fair comparisons of AIWP and NWP model output},
  author        = {Tilmann Gneiting and Tobias Biegert and Kristof Kraus and Eva-Maria Walz and Alexander I. Jordan and Sebastian Lerch},
  year          = {2025},
  eprint        = {2506.03744},
  archiveprefix = {arXiv},
  primaryclass  = {stat.AP},
  url           = {https://arxiv.org/abs/2506.03744}
}

@article{Hering_2011, title={Comparing Spatial Predictions}, volume={53}, ISSN={1537-2723}, url={http://dx.doi.org/10.1198/tech.2011.10136}, DOI={10.1198/tech.2011.10136}, number={4}, journal={Technometrics}, publisher={Informa UK Limited}, author={Hering, Amanda S. and Genton, Marc G.}, year={2011}, month=Nov, pages={414–425} }

@article{Hersbach_2000,
  title     = {Decomposition of the Continuous Ranked Probability Score for Ensemble Prediction Systems},
  volume    = {15},
  issn      = {1520-0434},
  url       = {http://dx.doi.org/10.1175/1520-0434(2000)015<0559:dotcrp>2.0.co;2},
  doi       = {10.1175/1520-0434(2000)015<0559:dotcrp>2.0.co;2},
  number    = {5},
  journal   = {Weather and Forecasting},
  publisher = {American Meteorological Society},
  author    = {Hersbach, Hans},
  year      = {2000},
  month     = oct,
  pages     = {559–570}
}

@article{hersbach2020era5,
  title     = {The ERA5 global reanalysis},
  author    = {Hersbach, Hans and Bell, Bill and Berrisford, Paul and Hirahara, Shoji and Hor{\'a}nyi, Andr{\'a}s and Mu{\~n}oz-Sabater, Joaqu{\'\i}n and Nicolas, Julien and Peubey, Carole and Radu, Raluca and Schepers, Dinand and others},
  journal   = {Quarterly journal of the royal meteorological society},
  volume    = {146},
  number    = {730},
  doi       = {https://doi.org/10.1002/qj.3803},
  pages     = {1999--2049},
  year      = {2020},
  publisher = {Wiley Online Library}
}

@article{Henzi_2021,
  title     = {Isotonic Distributional Regression},
  volume    = {83},
  issn      = {1467-9868},
  url       = {http://dx.doi.org/10.1111/rssb.12450},
  doi       = {10.1111/rssb.12450},
  number    = {5},
  journal   = {Journal of the Royal Statistical Society Series B: Statistical Methodology},
  publisher = {Oxford University Press (OUP)},
  author    = {Henzi, Alexander and Ziegel, Johanna F. and Gneiting, Tilmann},
  year      = {2021},
  month     = aug,
  pages     = {963–993}
}

@article{Ikeda_2013,
  title     = {Evaluation of Cold-Season Precipitation Forecasts Generated by the Hourly Updating High-Resolution Rapid Refresh Model},
  volume    = {28},
  issn      = {1520-0434},
  url       = {http://dx.doi.org/10.1175/waf-d-12-00085.1},
  doi       = {10.1175/waf-d-12-00085.1},
  number    = {4},
  journal   = {Weather and Forecasting},
  publisher = {American Meteorological Society},
  author    = {Ikeda, Kyoko and Steiner, Matthias and Pinto, James and Alexander, Curtis},
  year      = {2013},
  month     = jul,
  pages     = {921–939}
}

@misc{jin2024,
  title         = {WeatherReal: A Benchmark Based on In-Situ Observations for Evaluating Weather Models},
  author        = {Weixin Jin and Jonathan Weyn and Pengcheng Zhao and Siqi Xiang and Jiang Bian and Zuliang Fang and Haiyu Dong and Hongyu Sun and Kit Thambiratnam and Qi Zhang},
  year          = {2024},
  eprint        = {2409.09371},
  archiveprefix = {arXiv},
  primaryclass  = {physics.ao-ph},
  url           = {https://arxiv.org/abs/2409.09371}
}

@misc{keisler2022forecastingglobalweathergraph,
  title         = {Forecasting Global Weather with Graph Neural Networks},
  author        = {Ryan Keisler},
  year          = {2022},
  eprint        = {2202.07575},
  archiveprefix = {arXiv},
  primaryclass  = {physics.ao-ph},
  url           = {https://arxiv.org/abs/2202.07575}
}

@article{Lam_2023,
  title     = {Learning skillful medium-range global weather forecasting},
  volume    = {382},
  issn      = {1095-9203},
  url       = {http://dx.doi.org/10.1126/science.adi2336},
  doi       = {10.1126/science.adi2336},
  number    = {6677},
  journal   = {Science},
  publisher = {American Association for the Advancement of Science (AAAS)},
  author    = {Lam, Remi and Sanchez-Gonzalez, Alvaro and Willson, Matthew and Wirnsberger, Peter and Fortunato, Meire and Alet, Ferran and Ravuri, Suman and Ewalds, Timo and Eaton-Rosen, Zach and Hu, Weihua and Merose, Alexander and Hoyer, Stephan and Holland, George and Vinyals, Oriol and Stott, Jacklynn and Pritzel, Alexander and Mohamed, Shakir and Battaglia, Peter},
  year      = {2023},
  month     = dec,
  pages     = {1416–1421}
}

@article{Lang_2026, title={AIFS-CRPS: ensemble forecasting using a model trained with a loss function based on the continuous ranked probability score}, volume={2}, ISSN={3005-1460}, url={http://dx.doi.org/10.1038/s44387-026-00073-7}, DOI={10.1038/s44387-026-00073-7}, number={1}, journal={npj Artificial Intelligence}, publisher={Springer Science and Business Media LLC}, author={Lang, Simon and Alexe, Mihai and Clare, Mariana C. A. and Roberts, Christopher and Adewoyin, Rilwan and Ben Bouallègue, Zied and Chantry, Matthew and Dramsch, Jesper and Dueben, Peter D. and Hahner, Sara and Maciel, Pedro and Prieto-Nemesio, Ana and O’Brien, Cathal and Pinault, Florian and Polster, Jan and Raoult, Baudouin and Tietsche, Steffen and Leutbecher, Martin}, year={2026}, month={Feb} }

@misc{lang2024aifsecmwfsdatadriven,
      title={AIFS -- ECMWF's data-driven forecasting system}, 
      author={Simon Lang and Mihai Alexe and Matthew Chantry and Jesper Dramsch and Florian Pinault and Baudouin Raoult and Mariana C. A. Clare and Christian Lessig and Michael Maier-Gerber and Linus Magnusson and Zied Ben Bouallègue and Ana Prieto Nemesio and Peter D. Dueben and Andrew Brown and Florian Pappenberger and Florence Rabier},
      year={2024},
      eprint={2406.01465},
      archivePrefix={arXiv},
      primaryClass={physics.ao-ph},
      url={https://arxiv.org/abs/2406.01465}, 
}

@article{Lavers_2022,
  title     = {An evaluation of ERA5 precipitation for climate monitoring},
  volume    = {148},
  issn      = {1477-870X},
  url       = {http://dx.doi.org/10.1002/qj.4351},
  doi       = {10.1002/qj.4351},
  number    = {748},
  journal   = {Quarterly Journal of the Royal Meteorological Society},
  publisher = {Wiley},
  author    = {Lavers, David A. and Simmons, Adrian and Vamborg, Freja and Rodwell, Mark J.},
  year      = {2022},
  month     = aug,
  pages     = {3152–3165}
}

@article{Leeuwenburg_2024,
  title     = {scores: A Python package for verifying and evaluating models and predictions with xarray},
  volume    = {9},
  issn      = {2475-9066},
  url       = {http://dx.doi.org/10.21105/joss.06889},
  doi       = {10.21105/joss.06889},
  number    = {99},
  journal   = {Journal of Open Source Software},
  publisher = {The Open Journal},
  author    = {Leeuwenburg, Tennessee and Loveday, Nicholas and Ebert, Elizabeth E. and Cook, Harrison and Khanarmuei, Mohammadreza and Taggart, Robert J. and Ramanathan, Nikeeth and Carroll, Maree and Chong, Stephanie and Griffiths, Aidan and Sharples, John},
  year      = {2024},
  month     = jul,
  pages     = {6889}
}

@software{leeuwenburg_2026_18638494,
	author = {Leeuwenburg, Tennessee and Loveday, Nicholas and Ramanathan, Nikeeth and Chong, Stephanie and Taggart, Robert J. and Shrestha, Durga and Khanarmuei, Mohammadreza and Cook, Harrison and Bluett, Liam and Ebert, Elizabeth E. and Carroll, Maree and Trotta, Belinda and Sharples, John and Bishop, Sam and Squire, Dougal T. and Griffiths, Aidan and Pagano, Thomas C. and Fisher, A.J. and Mandelbaum, Taylor and Jinghan, Fu and Smith, Paul R. and Abellan, Esteban and Beunk, Jurian and Esperson, Felix and Smallwood, J. and Wu, Xiaoxi},
	doi = {10.5281/zenodo.18638494},
	month = feb,
	publisher = {Zenodo},
	title = {scores: Metrics for the verification, evaluation and optimisation of forecasts, predictions or models},
	url = {https://doi.org/10.5281/zenodo.18638494},
	version = {2.5.0},
	year = 2026}

@software{loveday_2026_code,
  author    = {Loveday, Nicholas},
  title     = {Evaluating Extreme Precipitation Forecasts: A Threshold-Weighted, Spatial Verification Approach for Comparing an AI Weather Prediction Model Against a High-Resolution NWP Model},
  year      = {2026},
  version   = {2},
  publisher = {Zenodo},
  doi       = {10.5281/zenodo.20438050},
  url       = {https://doi.org/10.5281/zenodo.20438050}
}

@dataset{loveday_2026_data,
  author    = {Loveday, Nicholas and Hertneky, Tracy},
  title     = {Data for: Evaluating Extreme Precipitation Forecasts: A Threshold-Weighted, Spatial Verification Approach for Comparing an AI Weather Prediction Model Against a High-Resolution NWP Model},
  year      = {2026},
  publisher = {Zenodo},
  doi       = {10.5281/zenodo.19672887},
  url       = {https://doi.org/10.5281/zenodo.19672887}
}

@article{Lerch_2017,
  title     = {Forecaster’s Dilemma: Extreme Events and Forecast Evaluation},
  volume    = {32},
  issn      = {0883-4237},
  url       = {http://dx.doi.org/10.1214/16-sts588},
  doi       = {10.1214/16-sts588},
  number    = {1},
  journal   = {Statistical Science},
  publisher = {Institute of Mathematical Statistics},
  author    = {Lerch, Sebastian and Thorarinsdottir, Thordis L. and Ravazzolo, Francesco and Gneiting, Tilmann},
  year      = {2017},
  month     = feb
}

@article{Loveday_2024,
  title     = {The Jive Verification System and Its Transformative Impact on Weather Forecasting Operations},
  volume    = {105},
  issn      = {1520-0477},
  url       = {http://dx.doi.org/10.1175/bams-d-23-0267.1},
  doi       = {10.1175/bams-d-23-0267.1},
  number    = {11},
  journal   = {Bulletin of the American Meteorological Society},
  publisher = {American Meteorological Society},
  author    = {Loveday, Nicholas and Griffiths, Deryn and Leeuwenburg, Tennessee and Taggart, Robert J. and Pagano, Thomas C. and Cheng, George and Plastow, Kevin and Ebert, Elizabeth E. and Templeton, Cassandra and Carroll, Maree and Khanarmuei, Mohammadreza and Nagpal, Isha},
  year      = {2024},
  month     = nov,
  pages     = {E2047–E2063}
}

@article{Mass_2002,
  title     = {Does Increasing Horizontal Resolution Produce More Skillful Forecasts?},
  volume    = {83},
  issn      = {1520-0477},
  url       = {http://dx.doi.org/10.1175/1520-0477(2002)083<0407:dihrpm>2.3.co;2},
  doi       = {10.1175/1520-0477(2002)083<0407:dihrpm>2.3.co;2},
  number    = {3},
  journal   = {Bulletin of the American Meteorological Society},
  publisher = {American Meteorological Society},
  author    = {Mass, Clifford F. and Ovens, David and Westrick, Ken and Colle, Brian A.},
  year      = {2002},
  month     = mar,
  pages     = {407–430}
}

@misc{mcgovern2026extremeweatherbenchframework,
      title={Extreme Weather Bench: A framework and benchmark for evaluation of high-impact weather}, 
      author={Amy McGovern and Taylor Mandelbaum and Daniel Rothenberg and Nicholas Loveday and Corey Potvin and Montgomery Flora and Linus Magnusson and Eric Gilleland and John Allen},
      year={2026},
      eprint={2605.01126},
      archivePrefix={arXiv},
      primaryClass={cs.LG},
      url={https://arxiv.org/abs/2605.01126}, 
}

@article{Matheson_1976,
  title     = {Scoring Rules for Continuous Probability Distributions},
  volume    = {22},
  issn      = {1526-5501},
  url       = {http://dx.doi.org/10.1287/mnsc.22.10.1087},
  doi       = {10.1287/mnsc.22.10.1087},
  number    = {10},
  journal   = {Management Science},
  publisher = {Institute for Operations Research and the Management Sciences (INFORMS)},
  author    = {Matheson, James E. and Winkler, Robert L.},
  year      = {1976},
  month     = jun,
  pages     = {1087–1096}
}

@article{Mittermaier_2014,
  title     = {A Strategy for Verifying Near-Convection-Resolving Model Forecasts at Observing Sites},
  volume    = {29},
  issn      = {1520-0434},
  url       = {http://dx.doi.org/10.1175/waf-d-12-00075.1},
  doi       = {10.1175/waf-d-12-00075.1},
  number    = {2},
  journal   = {Weather and Forecasting},
  publisher = {American Meteorological Society},
  author    = {Mittermaier, Marion P.},
  year      = {2014},
  month     = apr,
  pages     = {185–204}
}

@article{Mittermaier_2017,
  title     = {Ensemble versus Deterministic Performance at the Kilometer Scale},
  volume    = {32},
  issn      = {1520-0434},
  url       = {http://dx.doi.org/10.1175/waf-d-16-0164.1},
  doi       = {10.1175/waf-d-16-0164.1},
  number    = {5},
  journal   = {Weather and Forecasting},
  publisher = {American Meteorological Society},
  author    = {Mittermaier, M. P. and Csima, G.},
  year      = {2017},
  month     = sep,
  pages     = {1697–1709}
}

@techreport{morisseau2025object,
  title       = {Object-oriented verification of TC-Jasper rainfall forecasts: Machine learning},
  author      = {Morisseau, Hector and Zhu, Hongyan and Hudson, Debra and de Burgh-Day, Catherine},
  institution = {Bureau of Meteorology},
  year        = {2025},
  number      = {Bureau Research Report No. 106},
  url         = {http://www.bom.gov.au/research/publications/researchreports/BRR-106.pdf}
}

@article{Murphy_1989, title={Skill Scores and Correlation Coefficients in Model Verification}, volume={117}, ISSN={1520-0493}, url={http://dx.doi.org/10.1175/1520-0493(1989)117<0572:ssacci>2.0.co;2}, DOI={10.1175/1520-0493(1989)117<0572:ssacci>2.0.co;2}, number={3}, journal={Monthly Weather Review}, publisher={American Meteorological Society}, author={Murphy, Allan H. and Epstein, Edward S.}, year={1989}, month=Mar, pages={572–582} }

@article{Nipen_2026, title={Regional Data-Driven Weather Modeling with a Global Stretched Grid}, volume={5}, ISSN={2769-7525}, url={http://dx.doi.org/10.1175/aies-d-25-0001.1}, DOI={10.1175/aies-d-25-0001.1}, number={2}, journal={Artificial Intelligence for the Earth Systems}, publisher={American Meteorological Society}, author={Nipen, Thomas Nils and Haugen, Håvard Homleid and Ingstad, Magnus Sikora and Nordhagen, Even Marius and Salihi, Aram Farhad Shafiq and Tedesco, Paulina and Seierstad, Ivar Ambjørn and Kristiansen, Jørn and Lang, Simon and Alexe, Mihai and Dramsch, Jesper and Raoult, Baudouin and Mertes, Gert and Chantry, Matthew}, year={2026}, month=Apr }

@article{Olivetti_2024,
  title     = {Do data-driven models beat numerical models in forecasting weather extremes? A comparison of IFS HRES, Pangu-Weather, and GraphCast},
  volume    = {17},
  issn      = {1991-9603},
  url       = {http://dx.doi.org/10.5194/gmd-17-7915-2024},
  doi       = {10.5194/gmd-17-7915-2024},
  number    = {21},
  journal   = {Geoscientific Model Development},
  publisher = {Copernicus GmbH},
  author    = {Olivetti, Leonardo and Messori, Gabriele},
  year      = {2024},
  month     = nov,
  pages     = {7915–7962}
}

@article{Pagano_2024,
  title     = {Challenges of Operational Weather Forecast Verification and Evaluation},
  volume    = {105},
  issn      = {1520-0477},
  url       = {http://dx.doi.org/10.1175/bams-d-22-0257.1},
  doi       = {10.1175/bams-d-22-0257.1},
  number    = {4},
  journal   = {Bulletin of the American Meteorological Society},
  publisher = {American Meteorological Society},
  author    = {Pagano, Thomas C. and Casati, Barbara and Landman, Stephanie and Loveday, Nicholas and Taggart, Robert and Ebert, Elizabeth E. and Khanarmuei, Mohammadreza and Jensen, Tara L. and Mittermaier, Marion and Roberts, Helen and Willington, Steve and Roberts, Nigel and Sowko, Mike and Strassberg, Gordon and Kluepfel, Charles and Bullock, Timothy A. and Turner, David D. and Pappenberger, Florian and Osborne, Neal and Noble, Chris},
  year      = {2024},
  month     = apr,
  pages     = {E789–E802}
}

@article{pic2024properscoringrulesmultivariate,
  title     = {Proper scoring rules for multivariate probabilistic forecasts based on aggregation and transformation},
  volume    = {11},
  issn      = {2364-3587},
  url       = {http://dx.doi.org/10.5194/ascmo-11-23-2025},
  doi       = {10.5194/ascmo-11-23-2025},
  number    = {1},
  journal   = {Advances in Statistical Climatology, Meteorology and Oceanography},
  publisher = {Copernicus GmbH},
  author    = {Pic, Romain and Dombry, Clément and Naveau, Philippe and Taillardat, Maxime},
  year      = {2025},
  month     = mar,
  pages     = {23–58}
}

@article{Price_2024,
  title     = {Probabilistic weather forecasting with machine learning},
  volume    = {637},
  issn      = {1476-4687},
  url       = {http://dx.doi.org/10.1038/s41586-024-08252-9},
  doi       = {10.1038/s41586-024-08252-9},
  number    = {8044},
  journal   = {Nature},
  publisher = {Springer Science and Business Media LLC},
  author    = {Price, Ilan and Sanchez-Gonzalez, Alvaro and Alet, Ferran and Andersson, Tom R. and El-Kadi, Andrew and Masters, Dominic and Ewalds, Timo and Stott, Jacklynn and Mohamed, Shakir and Battaglia, Peter and Lam, Remi and Willson, Matthew},
  year      = {2024},
  month     = dec,
  pages     = {84–90}
}

@article{Radford_2025,
  title     = {Accelerating Community-Wide Evaluation of AI Models for Global Weather Prediction by Facilitating Access to Model Output},
  volume    = {106},
  issn      = {1520-0477},
  url       = {http://dx.doi.org/10.1175/bams-d-24-0057.1},
  doi       = {10.1175/bams-d-24-0057.1},
  number    = {1},
  journal   = {Bulletin of the American Meteorological Society},
  publisher = {American Meteorological Society},
  author    = {Radford, Jacob T. and Ebert-Uphoff, Imme and Stewart, Jebb Q. and Musgrave, Kate D. and DeMaria, Robert and Tourville, Natalie and Hilburn, Kyle},
  year      = {2025},
  month     = jan,
  pages     = {E68–E76}
}

@article{Radford_2025_comparison,
  title     = {A Comparison of AI Weather Prediction and Numerical Weather Prediction Models for 1–7-Day Precipitation Forecasts},
  issn      = {1520-0434},
  url       = {http://dx.doi.org/10.1175/waf-d-24-0081.1},
  doi       = {10.1175/waf-d-24-0081.1},
  journal   = {Weather and Forecasting},
  publisher = {American Meteorological Society},
  author    = {Radford, Jacob T. and Ebert-Uphoff, Imme and Stewart, Jebb Q.},
  year      = {2025},
  month     = mar
}

@article{Rasp_2024,
  title     = {WeatherBench 2: A Benchmark for the Next Generation of Data‐Driven Global Weather Models},
  volume    = {16},
  issn      = {1942-2466},
  url       = {http://dx.doi.org/10.1029/2023ms004019},
  doi       = {10.1029/2023ms004019},
  number    = {6},
  journal   = {Journal of Advances in Modeling Earth Systems},
  publisher = {American Geophysical Union (AGU)},
  author    = {Rasp, Stephan and Hoyer, Stephan and Merose, Alexander and Langmore, Ian and Battaglia, Peter and Russell, Tyler and Sanchez‐Gonzalez, Alvaro and Yang, Vivian and Carver, Rob and Agrawal, Shreya and Chantry, Matthew and Ben Bouallegue, Zied and Dueben, Peter and Bromberg, Carla and Sisk, Jared and Barrington, Luke and Bell, Aaron and Sha, Fei},
  year      = {2024},
  month     = jun
}

@article{Rodwell_2010,
  title     = {A new equitable score suitable for verifying precipitation in numerical weather prediction},
  volume    = {136},
  issn      = {1477-870X},
  url       = {http://dx.doi.org/10.1002/qj.656},
  doi       = {10.1002/qj.656},
  number    = {650},
  journal   = {Quarterly Journal of the Royal Meteorological Society},
  publisher = {Wiley},
  author    = {Rodwell, Mark J. and Richardson, David S. and Hewson, Tim D. and Haiden, Thomas},
  year      = {2010},
  month     = jul,
  pages     = {1344–1363}
}

@article{Schwartz_2017,
  title     = {Generating Probabilistic Forecasts from Convection-Allowing Ensembles Using Neighborhood Approaches: A Review and Recommendations},
  volume    = {145},
  issn      = {1520-0493},
  url       = {http://dx.doi.org/10.1175/mwr-d-16-0400.1},
  doi       = {10.1175/mwr-d-16-0400.1},
  number    = {9},
  journal   = {Monthly Weather Review},
  publisher = {American Meteorological Society},
  author    = {Schwartz, Craig S. and Sobash, Ryan A.},
  year      = {2017},
  month     = sep,
  pages     = {3397–3418}
}

@article{Siegert_2017, title={Simplifying and generalising Murphy’s Brier score decomposition}, volume={143}, ISSN={1477-870X}, url={http://dx.doi.org/10.1002/qj.2985}, DOI={10.1002/qj.2985}, number={703}, journal={Quarterly Journal of the Royal Meteorological Society}, publisher={Wiley}, author={Siegert, Stefan}, year={2017}, month=Jan, pages={1178–1183} }

@article{Stein_2022,
  title     = {Neighborhood-Based Ensemble Evaluation Using the CRPS},
  volume    = {150},
  issn      = {1520-0493},
  url       = {http://dx.doi.org/10.1175/mwr-d-21-0224.1},
  doi       = {10.1175/mwr-d-21-0224.1},
  number    = {8},
  journal   = {Monthly Weather Review},
  publisher = {American Meteorological Society},
  author    = {Stein, Joël and Stoop, Fabien},
  year      = {2022},
  month     = aug,
  pages     = {1901–1914}
}

@misc{subich2025fixingdoublepenaltydatadriven,
  title         = {Fixing the Double Penalty in Data-Driven Weather Forecasting Through a Modified Spherical Harmonic Loss Function},
  author        = {Christopher Subich and Syed Zahid Husain and Leo Separovic and Jing Yang},
  year          = {2025},
  eprint        = {2501.19374},
  archiveprefix = {arXiv},
  primaryclass  = {cs.LG},
  url           = {https://arxiv.org/abs/2501.19374}
}

@article{Sz_kely_2005,
  title     = {A new test for multivariate normality},
  volume    = {93},
  issn      = {0047-259X},
  url       = {http://dx.doi.org/10.1016/j.jmva.2003.12.002},
  doi       = {10.1016/j.jmva.2003.12.002},
  number    = {1},
  journal   = {Journal of Multivariate Analysis},
  publisher = {Elsevier BV},
  author    = {Székely, Gábor J. and Rizzo, Maria L.},
  year      = {2005},
  month     = mar,
  pages     = {58–80}
}

@article{Taggart_2021,
  title     = {Evaluation of point forecasts for extreme events using consistent scoring functions},
  volume    = {148},
  issn      = {1477-870X},
  url       = {http://dx.doi.org/10.1002/qj.4206},
  doi       = {10.1002/qj.4206},
  number    = {742},
  journal   = {Quarterly Journal of the Royal Meteorological Society},
  publisher = {Wiley},
  author    = {Taggart, Robert},
  year      = {2021},
  month     = nov,
  pages     = {306–320}
}

@article{Theis_2005, title={Probabilistic precipitation forecasts from a deterministic model: a pragmatic approach}, volume={12}, ISSN={1469-8080}, url={http://dx.doi.org/10.1017/s1350482705001763}, DOI={10.1017/s1350482705001763}, number={3}, journal={Meteorological Applications}, publisher={Wiley}, author={Theis, S. E. and Hense, A. and Damrath, U.}, year={2005}, month=sep, pages={257–268} }

@article{Tustison_2001,
  title     = {Scale issues in verification of precipitation forecasts},
  volume    = {106},
  issn      = {0148-0227},
  url       = {http://dx.doi.org/10.1029/2001jd900066},
  doi       = {10.1029/2001jd900066},
  number    = {D11},
  journal   = {Journal of Geophysical Research: Atmospheres},
  publisher = {American Geophysical Union (AGU)},
  author    = {Tustison, Ben and Harris, Daniel and Foufoula‐Georgiou, Efi},
  year      = {2001},
  month     = jun,
  pages     = {11775–11784}
}

@article{Walz_2024,
  title     = {Easy Uncertainty Quantification (EasyUQ): Generating Predictive Distributions from Single-Valued Model Output},
  volume    = {66},
  issn      = {1095-7200},
  url       = {http://dx.doi.org/10.1137/22m1541915},
  doi       = {10.1137/22m1541915},
  number    = {1},
  journal   = {SIAM Review},
  publisher = {Society for Industrial & Applied Mathematics (SIAM)},
  author    = {Walz, Eva-Maria and Henzi, Alexander and Ziegel, Johanna and Gneiting, Tilmann},
  year      = {2024},
  month     = feb,
  pages     = {91–122}
}

@article{Winkler_1968,
  title     = {“Good” Probability Assessors},
  volume    = {7},
  issn      = {0021-8952},
  url       = {http://dx.doi.org/10.1175/1520-0450(1968)007<0751:pa>2.0.co;2},
  doi       = {10.1175/1520-0450(1968)007<0751:pa>2.0.co;2},
  number    = {5},
  journal   = {Journal of Applied Meteorology},
  publisher = {American Meteorological Society},
  author    = {Winkler, Robert L. and Murphy, Allan H.},
  year      = {1968},
  month     = oct,
  pages     = {751–758}
}

@article{wessel2024improvingprobabilisticforecastsextreme,
  title     = {Improving probabilistic forecasts of extreme wind speeds by training statistical post-processing models with weighted scoring rules},
  issn      = {1520-0493},
  url       = {http://dx.doi.org/10.1175/mwr-d-24-0151.1},
  doi       = {10.1175/mwr-d-24-0151.1},
  journal   = {Monthly Weather Review},
  publisher = {American Meteorological Society},
  author    = {Wessel, Jakob Benjamin and Ferro, Christopher A. T. and Evans, Gavin R. and Kwasniok, Frank},
  year      = {2025},
  month     = apr
}

@article{Zhang_2026, title={Physics-based models outperform AI weather forecasts of record-breaking extremes}, volume={12}, ISSN={2375-2548}, url={http://dx.doi.org/10.1126/sciadv.aec1433}, DOI={10.1126/sciadv.aec1433}, number={18}, journal={Science Advances}, publisher={American Association for the Advancement of Science (AAAS)}, author={Zhang, Zhongwei and Fischer, Erich and Zscheischler, Jakob and Engelke, Sebastian}, year={2026}, month={May}}

@techreport{NWS_ASOS,
  author      = {NWS},
  title       = {Automated Surface Observing System (ASOS) user's guide},
  institution = {NOAA},
  pages       = {66},
  year        = {1998}
}

@techreport{WMO1994,
  author      = {WMO},
  title       = {Guide to Hydrological Practices},
  institution = {World Meteorological Organization},
  number      = {168},
  year        = {1994}
}

\end{document}